\documentclass [12pt]{article}
\usepackage{amsmath,amssymb,cite}
\usepackage{appendix}
\usepackage{feynmp}
\usepackage[vcentermath,enableskew]{youngtab}
\usepackage{tabularx}
\usepackage[english]{babel}
\usepackage[dvips]{graphicx}
\usepackage{indentfirst}
\usepackage{epsfig}
\usepackage{physics}
\usepackage{pdfpages}
\usepackage{cases}
\usepackage{appendix}
\usepackage{tabularx}
\usepackage[english]{babel}
\usepackage[dvips]{graphicx}
\usepackage{indentfirst}
\usepackage{epsfig}
\usepackage{slashed}
\usepackage{fancyhdr}
\usepackage{cases}
\usepackage{tcolorbox}
\usepackage{amsmath,amssymb,cite}
\usepackage{amsfonts, geometry, hyperref}
\usepackage{booktabs,longtable}
\usepackage{appendix}
\usepackage{feynmp}
\usepackage{float}
\usepackage{subfigure}
\usepackage[vcentermath,enableskew]{youngtab}
\usepackage{tabularx}
\usepackage[english]{babel}
\usepackage{graphicx}
\usepackage{indentfirst}
\usepackage{epsfig}
\usepackage{color}
\usepackage{epstopdf}
\usepackage[]{hyperref}
\usepackage[section]{placeins}
\usepackage[stable]{footmisc}
\usepackage{appendix}
\usepackage{adjustbox}
\usepackage{tabularx}
\usepackage[english]{babel}
\usepackage{graphicx}
\usepackage{indentfirst}
\usepackage{epsfig}
\usepackage{slashed}
\usepackage{fancyhdr}
\usepackage{url}
\usepackage{pgfplots}
\usepackage{amsmath}

\pgfplotsset{compat=1.18}
\numberwithin{equation}{section}

\makeatletter
\patchcmd{\beamer@sectionintoc}
  {\vfill}
  {\vskip\itemsep}
  {}
  {}

\long\def\beamer@section[#1]#2{%
  \beamer@savemode%
  \mode<all>%
  \ifbeamer@inlecture
    \refstepcounter{section}%
    \beamer@ifempty{#2}%
    {\long\def\secname{#1}\long\def\lastsection{#1}}%
    {\global\advance\beamer@tocsectionnumber by 1\relax%
      \long\def\secname{#2}%
      \long\def\lastsection{#1}%
      \addtocontents{toc}{\protect\beamer@sectionintoc{\the\c@section}{#2\hfill\the\c@page}{\the\c@page}{\the\c@part}%
        {\the\beamer@tocsectionnumber}}}%
    {\let\\=\relax\xdef\sectionlink{{Navigation\the\c@page}{\noexpand\secname}}}%
    \beamer@tempcount=\c@page\advance\beamer@tempcount by -1%
    \beamer@ifempty{#1}{}{%
      \addtocontents{nav}{\protect\headcommand{\protect\sectionentry{\the\c@section}{#1}{\the\c@page}{\secname}{\the\c@part}}}%
      \addtocontents{nav}{\protect\headcommand{\protect\beamer@sectionpages{\the\beamer@sectionstartpage}{\the\beamer@tempcount}}}%
      \addtocontents{nav}{\protect\headcommand{\protect\beamer@subsectionpages{\the\beamer@subsectionstartpage}{\the\beamer@tempcount}}}%
    }%
    \beamer@sectionstartpage=\c@page%
    \beamer@subsectionstartpage=\c@page%
    \def\insertsection{\expandafter\hyperlink\sectionlink}%
    \def\insertsubsection{}%
    \def\insertsubsubsection{}%
    \def\insertsectionhead{\hyperlink{Navigation\the\c@page}{#1}}%
    \def\insertsubsectionhead{}%
    \def\insertsubsubsectionhead{}%
    \def\lastsubsection{}%
    \Hy@writebookmark{\the\c@section}{\secname}{Outline\the\c@part.\the\c@section}{2}{toc}%
    \hyper@anchorstart{Outline\the\c@part.\the\c@section}\hyper@anchorend%
    \beamer@ifempty{#2}{\beamer@atbeginsections}{\beamer@atbeginsection}%
  \fi%
  \beamer@resumemode}%

\def\beamer@subsection[#1]#2{%
  \beamer@savemode%
  \mode<all>%
  \ifbeamer@inlecture%
    \refstepcounter{subsection}%
    \beamer@ifempty{#2}{\long\def\subsecname{#1}\long\def\lastsubsection{#1}}
    {%
      \long\def\subsecname{#2}%
      \long\def\lastsubsection{#1}%
      \addtocontents{toc}{\protect\beamer@subsectionintoc{\the\c@section}{\the\c@subsection}{#2\hfill\the\c@page}{\the\c@page}{\the\c@part}{\the\beamer@tocsectionnumber}}%
    }%
    \beamer@tempcount=\c@page\advance\beamer@tempcount by -1%
    \addtocontents{nav}{%
      \protect\headcommand{\protect\beamer@subsectionentry{\the\c@part}{\the\c@section}{\the\c@subsection}{\the\c@page}{\lastsubsection}}%
      \protect\headcommand{\protect\beamer@subsectionpages{\the\beamer@subsectionstartpage}{\the\beamer@tempcount}}%
    }%
    \beamer@subsectionstartpage=\c@page%
    \edef\subsectionlink{{Navigation\the\c@page}{\noexpand\subsecname}}%
    \def\insertsubsection{\expandafter\hyperlink\subsectionlink}%
    \def\insertsubsubsection{}%
    \def\insertsubsectionhead{\hyperlink{Navigation\the\c@page}{#1}}%
    \def\insertsubsubsectionhead{}%
    \Hy@writebookmark{\the\c@subsection}{#2}{Outline\the\c@part.\the\c@section.\the\c@subsection.\the\c@page}{3}{toc}%
    \hyper@anchorstart{Outline\the\c@part.\the\c@section.\the\c@subsection.\the\c@page}\hyper@anchorend%
    \beamer@ifempty{#2}{\beamer@atbeginsubsections}{\beamer@atbeginsubsection}%
  \fi%
  \beamer@resumemode}

\makeatother

\begin{document}

\title{Gram--Wishart--Stiefel Formulation of the \(N=2\),
Large--\(d\) Gauge Theory in 1D:
Towards Emergent Grassmannian Geometry}

\author{
Badis Ydri\\[2mm]
Department of Physics, Badji Mokhtar Annaba University, Algeria\\
}

\maketitle

\begin{abstract}
  We develop the Gram/Wishart/Stiefel formulation of the \(N=2\), large--\(d\) planar endpoint theory of lattice BFSS/BMN matrix quantum mechanics on the lattice. The endpoint variables become rank--two Wishart eigenvalues and relative Stiefel angles, expressing the holonomy invariants \(A\), \(B\), and \(R^2=A^2+B^2\) in radial and angular Gram data. The large-\(R\) aligned contribution \(-A\) is absorbed into the Gaussian sector, producing the shifted mass \((\alpha_\Lambda)_{\rm eff}=\alpha_\Lambda-1/2\). The exact \(O(2)\) angular integral encodes this sector in a rank--two Bessel kernel. The pure \(-A\) theory, exactly solvable in Cartesian variables, fixes the leading Bessel/HCIZ structure: its exponential selects alignment, while its prefactor removes the spurious doubled Wishart entropy.

  We then apply this structure to the transverse \(B\)-type expansion and its non-polynomial toy completion. Finite polynomial truncations lead to an apparent large--\(d\) perturbativity bound incompatible with the continuum limit, but this bound is shown to be an artifact of truncation. After summing the local transverse completion and balancing the compensating \(+A\) term, the Wishart saddle is recovered with the physical shifted mass. The resulting continuum behavior reproduces the universal \(-2d\) contribution of the \(D_\Lambda\)-channel, while the genuinely anisotropic \(\beta_\Lambda\)-channel lies outside the scope of a pure transverse \(B\)-type description.

  Finally, we outline the emergent Grassmannian geometry encoded
by the two endpoint planes. Their relative angular fluctuations
form a \(2(d-2)\)-dimensional sector, whose entropy interpretation
is developed in the sequel. For \(d\ge4\), the two planes generically span a
relational four-dimensional subspace of \(\mathbb R^d\).
\end{abstract}

\bigskip
{\small
\par}

\newpage
\tableofcontents

\newpage

\section{Introductory remarks}
\subsection{Generalities}

\subsubsection{BFSS/BMN systems}

\medskip
\noindent
A basic class of matrix quantum mechanical models arises from the dimensional
reduction of supersymmetric Yang--Mills theory to one time dimension~\cite{Brink:1976bc}.
In particular, reducing \((d+1)\)-dimensional \(\mathcal N=1\) super Yang--Mills theory
yields a gauged quantum mechanics whose degrees of freedom are adjoint \(N\times N\) matrix
coordinates \(X_a(t)\), with Euclidean bosonic action

\begin{eqnarray}
S_{\rm BFSS,B}^{\rm E}
=
\frac{1}{g^2}
\int_{0}^{\beta}\! dt~ 
\Tr\!\left[
\frac{1}{2}(D_t X_a)^2
-
\frac{1}{4}[X_a,X_b]^2
\right]
+
\text{fermionic terms},
\qquad
D_t=\partial_t-i[A_t,\cdot].
\label{BFSSIntro}
\end{eqnarray}
The allowed supersymmetric reductions are restricted by the Fierz identity
analysis of Baake, Reinicke and Rittenberg~\cite{Baake:1984ie}, giving
\begin{eqnarray}
D_{\rm YM}=d+1=10,6,4,3,2.
\end{eqnarray}
Here \(d\) is the number of bosonic matrix coordinates in the reduced quantum
mechanics. The corresponding supermembrane or M-theory dimensions are shifted by
one further unit,
\begin{eqnarray}
D_{\rm M}=d+2=11,7,5,4,3.
\end{eqnarray}

\medskip
\noindent
The holographic regime is the usual planar limit~\cite{tHooft1974},
\begin{eqnarray}
N\to\infty,
\qquad
g^2\to0,
\qquad
\lambda=g^2N
\quad \hbox{fixed}.
\end{eqnarray}
This large-\(N\) limit is the natural setting for holography~\cite{tHooft1993,Susskind1995}.

\medskip
\noindent
The central example is the \(d=9\) case, namely the
Banks--Fischler--Shenker--Susskind BFSS\(_{10}\) model, or M-(atrix)
theory~\cite{BanksFischlerShenkerSusskind1997}. In the large-\(N\) limit, it
gives a non-perturbative formulation of M-theory in the infinite-momentum frame.
Equivalently, it describes the low-energy dynamics of \(N\) coincident
D\(0\)-branes~\cite{Witten1996}. At strong coupling, this matrix quantum
mechanics is dual to type IIA supergravity on the black 0-brane background
~\cite{Itzhaki1998}, the gravitational description of a D\(0\)-brane bound
state in type IIA string theory~\cite{Polchinski1995}.

\medskip
\noindent
This picture is tied to the eleven-dimensional origin of type IIA supergravity:
compactifying eleven-dimensional supergravity~\cite{Cremmer1978} on a circle
gives type IIA theory~\cite{Witten1995}, while reducing the M-wave solution gives
the ten-dimensional non-extremal black 0-brane
~\cite{Hyakutake:2014maa,Hyakutake:2006aq}. The same BFSS model is therefore
also connected to light-cone supermembranes
~\cite{Hoppe1982,Hoppe1988,deWitHoppeNicolai1988} and to superparticle dynamics
in maximally supersymmetric pp-wave backgrounds
~\cite{Kowalski-Glikman:1984qtj,Blau:2001ne}.

In this interpretation, the matrices \(X_a\) are noncommutative coordinates for
the D\(0\)-brane system. Their diagonal entries describe the transverse positions
of the individual branes, while the off-diagonal entries encode the open-string
modes stretching between different branes. When the branes coincide, these
off-diagonal modes become light and generate the non-Abelian gauge dynamics; see
for example~\cite{Azeyanagi2009} and the pedagogical accounts
~\cite{Zwiebach2009,Becker2006}.

\medskip
\noindent
Maldacena's gauge/gravity conjecture~\cite{Maldacena1999,Gubser1998,Witten1998}
then identifies the strongly coupled large-\(N\) matrix quantum mechanics, in
the appropriate decoupling limit, with weakly coupled type II string theory on
the black 0-brane background. Since the gauge theory admits a nonperturbative
lattice regularization~\cite{Wilson:1974sk}, this provides a non-perturbative framework for studying quantum gravity and
black-hole thermodynamics through matrix quantum mechanics.

\medskip
\noindent
This correspondence has been tested extensively by Monte Carlo simulations
~\cite{Catterall2008,Anagnostopoulos2008,Hanada2014,Hanada2016b,Filev:2015hiaF}
and by analytic methods~\cite{Kabat2001,Hanada2009,Hyakutake2014}; see
~\cite{Hanada2016} for a pedagogical review. The black 0-brane system therefore
provides one of the most concrete testbeds for gauge/gravity duality, connecting
matrix quantum mechanics to quantum black holes.

\medskip
\noindent
The lower-dimensional BFSS\(_{d+1}\) models with
\begin{eqnarray}
D_{\rm YM}=d+1=6,4,3,2
\end{eqnarray}
may be viewed as lower-dimensional analogues of BFSS\(_{10}\). They retain the
same basic structures---gauge holonomy, adjoint matrix dynamics, eigenvalue
interactions, and possible deconfinement behavior---while being more
tractable analytically and numerically. They also provide lower-dimensional
settings for light-cone matrix dynamics and noncritical holographic sectors.

\medskip
\noindent
A second important class is obtained by adding supersymmetric mass deformations.
The prototype is the BMN plane-wave matrix model~\cite{BerensteinMaldacenaNastase2002},
and the possible massive super Yang--Mills quantum mechanics were classified in
~\cite{Kim:2006,Park:2005}. At the bosonic level one may write schematically
\begin{eqnarray}
S_{\rm BMN,B}^{\rm E}
=
S_{\rm BFSS,B}^{\rm E}
+
\frac{1}{g^2}
\int_{0}^{\beta}\! dt~
\Tr\!\left[
\mu_1 X_a^2
+
\mu_2 \epsilon_{ijk}X_iX_jX_k
\right]
+
\text{fermionic terms}.
\label{BMNIntro}
\end{eqnarray}
The quadratic mass term lifts the flat directions, while the Myers interaction
~\cite{Myers} selects an \(SO(3)\) sector. The maximally supersymmetric cases
include \(d=2,3,5,9\), corresponding to BMN\(_{3,4,6,10}\), while the special
\(d=1\) case is discussed in~\cite{Park:2005,Ydri2025}. These deformed models
describe pp-wave rather than flat backgrounds. In the maximally supersymmetric case, their half-BPS sectors are related to LLM
bubbling geometries~\cite{Lin:2004nb}, while their thermal dynamics has been
studied numerically, for example in~\cite{Asano:2018nol,Asano:2020yry}.

In summary, each BFSS\(_{d+1}\) model admits a corresponding BMN deformation preserving maximal supersymmetry. These deformations describe supermembranes and superparticles in maximally supersymmetric pp-wave backgrounds. The corresponding classification is summarized in Table~\ref{so3}.

\begin{table}[h]
\centering
\begin{tabular}{@{}lllll@{}}
\toprule
\textbf{Model} & \(D_{\rm YM}\) & Splitting of \(SO(D_{\rm YM}-1)\) & Superalgebra & Deformation parameter \\
\midrule
\(\mathcal N=16\) & 10 & \(SO(6)\times SO(3)\) & \(\mathfrak{su}(2|4)\) & \(\mu\) \\
\(\mathcal N=8\) type I & 6 & \(SO(3)\times SO(2)\) & \(\mathfrak{su}(2|2)\) & \(\mu\) \\
\(\mathcal N=8\) type II & 6 & \(SO(4)\) & \(\mathfrak{su}(2|1)\oplus\mathfrak{su}(2|1)\) & \(\mu\) \\
\(\mathcal N=4\) type I & 4 & \(SO(3)\) & \(\mathfrak{su}(2|1)\) & \(\mu_1,\mu_2\) \\
\(\mathcal N=4\) type II & 4 & \(SO(2)\) & \(\mathrm{Clifford}_{4}(\mathbb R)\) & \(\mu\) \\
\(\mathcal N=2\) & 3 & \(SO(2)\) & \(\mathrm{Clifford}_{2}(\mathbb R)\) & \(\mu\) \\
\(\mathcal N=1+1\) & 2 & \(SO(1,2)\) & \(\mathfrak{osp}(1|2,\mathbb R)\) & \(\Lambda(t),\rho(t)\) \\
\bottomrule
\end{tabular}
\caption{Classification of massive supersymmetric Yang--Mills quantum mechanics models and their deformation parameters.}
\label{so3}
\end{table}

\subsubsection{Large--\(d\) saddle and Molien--Weyl integrals}

\medskip
\noindent
A useful solvable limit of BFSS/BMN matrix quantum mechanics is obtained in the Gaussian, or large-mass, regime. In this limit the theory reduces to a supersymmetric gauged matrix harmonic oscillator, whose singlet-sector partition function can be written as a Molien--Weyl integral~\cite{OConnor:2023mss,OConnor:2024udv}. Mathematically, these integrals compute Hilbert series of invariant operators; see, for example,~\cite{CoxLittleOShea2005}.

\medskip
\noindent
The same structure also arises dynamically in the large--\(d\) limit of the BFSS systems \eqref{BFSSIntro}. In this limit the commutator interaction becomes self-consistently Gaussian, producing an effective mass \(s\sim d^{1/3}\)~\cite{Mandal:2009vzN,Mandal:2011hbN,Kabat:2000zv,Kabat:2001ve}. For the BMN systems \eqref{BMNIntro}, the corresponding double-scaling regime was identified in~\cite{Ydri:2026gsy}: one takes both the BMN deformation mass \(m\) and the number of matrices \(d\) large, while keeping
\begin{eqnarray}
\hat m=\kappa^{2/3}
\equiv
\frac{m}{d^{2/3}}
\end{eqnarray}
fixed.

\medskip
\noindent
The resulting effective theory is the gauged matrix harmonic oscillator
\begin{eqnarray}
S_{\rm MHO}[X;\theta]
=
N\int_{0}^{\beta}\! dt\,
\mathrm{Tr}\!\left[
\frac{1}{2}(D_tX_a)^2
+
\frac{s^2}{2}X_a^2
\right]
+
\text{fermionic terms},
\label{MHO_action_for_Ward}
\end{eqnarray}
with the self-consistent gap equation
\begin{eqnarray}
s^2=m+k_0,
\qquad
s^3-ms=d.
\end{eqnarray}
The corresponding confinement/deconfinement transition is governed by the holonomy effective action~\cite{Kawahara:2007fnF,Aharony:2003sxF,Aharony:2004igF,AlvarezGaume:2005fvF,Gross:1980heF,Wadia:1980cpF}, with critical temperature
\begin{eqnarray}
T_c(\kappa)
=
\frac{s(\kappa)}{\log d}
=
\frac{d^{1/3}}{\log d}
\left(
\kappa^{1/3}
+
\frac{1}{2\kappa^{2/3}}
+\cdots
\right).
\end{eqnarray}
Thus the transition is pushed to parametrically higher temperature as \(d\to\infty\).

\begin{figure}[htbp]
\begin{center}
   \includegraphics[width=10cm,angle=-0,page=3]{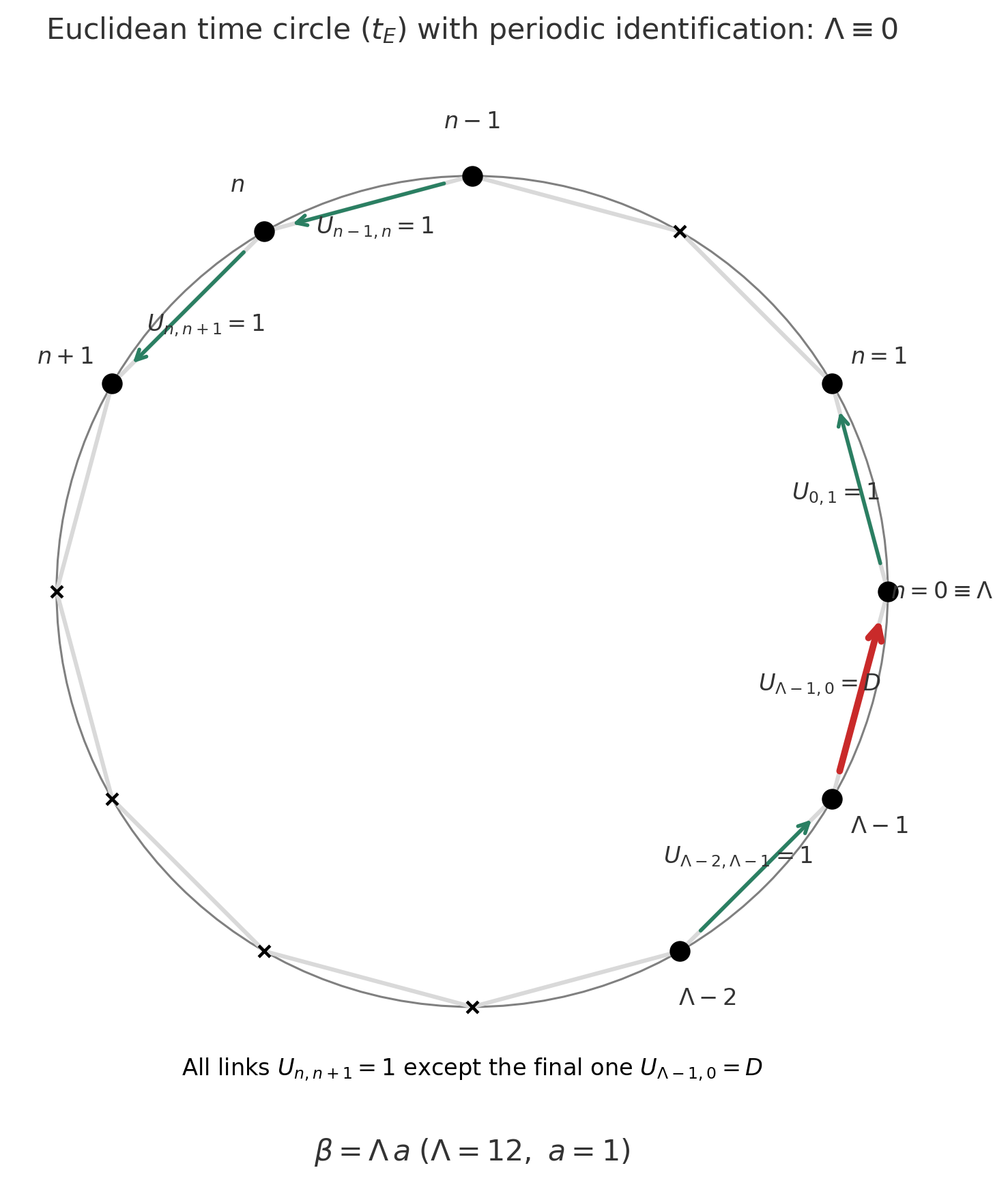}
\end{center}
\caption{The static diagonal (Polyakov) gauge.}\label{GF}
\end{figure}

\medskip
\noindent
On the lattice, the thermal circle is discretized with spacing \(a=\beta/\Lambda\), and the gauge field is represented by link variables \(U_{n,n+1}\). Using gauge invariance and Haar-measure invariance, one may work in the static Polyakov gauge~\cite{Filev:2015hiaF}, where all links are gauged to unity except the closing link. See Fig.~\ref{GF}. The gauge field is then reduced to a single holonomy
\begin{eqnarray}
g=\mathcal{P}\exp\!\left(i\int_0^\beta dt\,A_t\right)\in U(N).
\end{eqnarray}
After integrating out the Gaussian matrix coordinates, the partition function becomes a single group integral. Since the \(d\) matrices factorize, the normal-ordered bosonic partition function is
\begin{eqnarray}
Z_{N,d}(x)
=
\int d\mu(g)\;
\frac{1}{\Big[{\bf det}\big(1-x\,g\otimes g^{-1}\big)\Big]^d},
\qquad
x=e^{-\beta s}.
\label{eq:MW_BFSSd}
\end{eqnarray}

\medskip
\noindent
Diagonalizing \(g=\mathrm{diag}(z_1,\ldots,z_N)\) gives the standard Molien--Weyl form
\begin{eqnarray}
Z_{N,d}(x)
=
\frac{1}{N!}
\oint
\prod_{i=1}^{N}\frac{dz_i}{2\pi i z_i}\;
\Delta(z)\Delta(z^{-1})
\prod_{i,j=1}^{N}
\frac{1}{\big(1-x\,z_i z_j^{-1}\big)^d}.
\label{eq:MW_BFSSd_explicit}
\end{eqnarray}
Equivalently, in the notation used for bosonic Molien--Weyl factors,
\begin{eqnarray}
Z_{N,d}(x)
&=&
\frac{1}{N!}
\frac{1}{(1-x_b)^{n_bN}}
\oint
\prod_{i=1}^{N}\frac{dz_i}{2\pi i z_i}\;
\Delta_A(-1,z)\,
\frac{1}{\Delta_B^{n_b}(-x_b,z)},\nonumber\\
&&x_b=e^{-\beta m_b}\equiv x,\quad
m_b\equiv s,\quad
n_b\equiv d.
\label{eq:MW_BFSSd_explicit1}
\end{eqnarray}
Here \(1/N!\) is the Weyl-group factor in the Haar measure, and the determinants \(\Delta_A\) and \(\Delta_B\) are built from
\begin{eqnarray}
\Delta(x,z)
=
\prod_{i<j}
\left(1+x\frac{z_i}{z_j}\right)
\prod_{i<j}
\left(1+x\frac{z_j}{z_i}\right).
\end{eqnarray}

\medskip
\noindent
For \(U(N)\), the adjoint contains \(N\) zero weights, producing the factor \((1-x_b)^{-dN}\). Since
\(\mathfrak u(N)=\mathfrak{su}(N)\oplus\mathfrak u(1)\), one zero-weight factor belongs to the decoupled \(U(1)\) sector. Passing to the \(SU(N)\) adjoint therefore removes this factor:
\begin{eqnarray}
Z^{SU(N)}_{N,d}(x)
=
(1-x_b)^d\,Z^{U(N)}_{N,d}(x),
\end{eqnarray}
or equivalently
\begin{eqnarray}
Z_{N,d}(x)
=
\frac{1}{N!}
\frac{1}{(1-x_b)^{n_b(N-1)}}
\oint
\prod_{i=1}^{N}\frac{dz_i}{2\pi i z_i}\;
\Delta_A(-1,z)\,
\frac{1}{\Delta_B^{n_b}(-x_b,z)}.
\end{eqnarray}
The full supersymmetric versions of these gauged matrix harmonic oscillator models and their Molien--Weyl integrals are given in~\cite{Ydri:2026gsy,Ydri:2026smo}.

\subsection{Goal: towards matrix quantum mechanics and matrix quantum gravity}

\medskip
\noindent
This work is part of a broader program on gauge theory in one dimension with an
arbitrary number \(d\) of noncommuting matrix coordinates. The basic objects are
the BFSS\(_{d+1}\) matrix quantum mechanical models and their mass-deformed
BMN\(_{d+1}\) extensions. Our approach combines large-\(d\) Gaussian reduction,
Molien--Weyl singlet projection, endpoint/Wishart formulations, and Monte Carlo
methods, with the aim of clarifying the structure of matrix quantum mechanics
and its possible interpretations as matrix quantum gravity.

\medskip
\noindent
Here, quantum gravity may refer either to the gauge/gravity duality approach and its quantum black-hole dynamics, or to the noncommutative-geometry/matrix-model approach to emergent geometry and gravity \cite{Ydri:2022ueu,Ydri:2021cam,Ydri:2020fry}.

\medskip
\noindent
The program is guided by two working assumptions. The first is that matrix
quantum mechanics is the more fundamental dynamical framework, while
zero-dimensional matrix models, including the IKKT model~\cite{Ishibashi:1996xs},
may be viewed as reductions, limits, or approximations of an underlying
one-dimensional matrix quantum theory. The second is the ``unreasonable
effectiveness'' of Gaussian structures in matrix quantum mechanics: even before
the full interacting theory is restored, the Gaussian reduction already captures
a substantial part of the nontrivial quantum dynamics.

\medskip
\noindent
The broader project is organized around several connected themes~\cite{Ydri:2026gsy,Ydri:2026smo,YdriUnpublished0,Ydri:2026rke,YdriUnpublished2,YdriUnpublished3} and \cite{Ydri2025}:
the large-\(d\) limit of BFSS/BMN systems; Molien--Weyl singlet counting and
BFSS\(_2\) factorization; endpoint formulations of Gaussian matrix quantum
mechanics and their Wishart/Stiefel geometry; Monte Carlo studies of
supersymmetric BFSS\(_3\)/BMN\(_3\); and the relation of BFSS\(_2\)/BMN\(_2\)
to noncommutative AdS\(_2\)/dS\(_2\) geometry and emergent or latent geometry.

\medskip
\noindent
For related recent developments, see also the preprints
\cite{OConnor:2026zlf,Lei:2026fep}.

\medskip
\noindent
The present paper focuses on the Gram/Wishart/Stiefel formulation of the
\(N=2\), large-\(d\) planar endpoint theory. Its purpose is to understand how the
holonomy dynamics, the shifted Gaussian structure, and the universal
\(D_\Lambda\)-channel contribution are encoded in the radial Wishart variables
and the relative Stiefel angular sector.

\subsection{Organization of the paper}

\medskip
\noindent
This paper is organized as follows. Section~\ref{section2} recalls the planar endpoint model,
its constrained saddle, and the aligned \(-A\) contribution.
It also introduces the non-polynomial toy model, its
nontrivial continuum \(D_\Lambda\)-channel, and the completed
transverse potential.

\medskip
\noindent
Section~\ref{section4} develops the Gram/Wishart/Stiefel
formulation, separating the radial eigenvalues from the
relative angular variables. It introduces the weak--minor
approximation and treats the planar angular integral,
leading to the two-branch Bessel kernel and its aligned
large-argument behavior.

\medskip
\noindent
Section~\ref{section6} examines the pure \(-A\) theory.
Comparison with the exact Cartesian Gaussian integral
motivates the angular-kernel ansatz and identifies the
prefactor needed to remove the apparent doubled Wishart
entropy and recover the leading continuum coefficient.

\medskip
\noindent
Section~\ref{section7} extends this construction to the
transverse theory through a Bessel/HCIZ factorization ansatz.
The resulting shifted Wishart saddle exposes a second
obstruction: termwise perturbative control of the transverse
interaction is incompatible with the required continuum
scaling.

\medskip
\noindent
Section~\ref{section8} addresses this obstruction through
the summed local completion. The flat cancellation region
and the balanced compensating split recover the Wishart
saddle and the universal large-\(d\) \(D_\Lambda\)-channel
coefficient \(-2d\), without the apparent perturbative bound.

\medskip
\noindent
Finally, Section~\ref{conclusion} summarizes the calculation,
distinguishes the \(D_\Lambda\)- and \(\beta_\Lambda\)-channels,
and outlines the connection to emergent Grassmannian
geometry developed in the sequel \cite{YdriGrassmannian}. Technical derivations
omitted from this shortened presentation remain available
in the first arXiv version~\cite{YdriGramV1}.

\section{The planar boundary model}\label{section2}

\subsection{Transverse expansion of the planar action around its constrained saddle}

\medskip
\noindent 
The starting point is the endpoint form of the path integral for the
\emph{\(N=2\), large--\(d\) BFSS/BMN matrix quantum mechanics on the lattice}.

\medskip
\noindent
After integrating out the bulk fluctuations along the thermal circle,
the path integral reduces to a boundary theory involving only the two endpoints
\(n=0\) and \(n=\Lambda\). By gauge fixing to the static Polyakov gauge, the two endpoints are still
connected by the holonomy. The gauge field is then integrated out, followed by
the longitudinal endpoint variables.

\medskip
\noindent 
The remaining degrees of freedom are therefore transverse two-dimensional vectors. More precisely, for each matrix
direction \(a=1,\ldots,d\), one obtains two endpoint vectors
\begin{eqnarray}
V_a=(V_a^1,V_a^2),
\qquad
W_a=(W_a^1,W_a^2),
\qquad
a=1,\ldots,d,
\end{eqnarray}
where \(V_a\) and \(W_a\) encode the initial and final transverse endpoint
configurations of the \(a\)-th coordinate matrix. Thus the endpoint theory is
described by two sets of \(d\) two-dimensional vectors,
\begin{eqnarray}
\{V_a\}_{a=1}^{d},
\qquad
\{W_a\}_{a=1}^{d}.
\end{eqnarray}

\medskip
\noindent
In terms of these variables, the planar endpoint action takes the form
\begin{eqnarray}
S_{N=2,\rm eff}^{\rm BFSS_{d+1}}=
m\sum_{a=1}^{d}
\Big(
V_a^2+W_a^2
\Big)-2\beta_\Lambda A
+
V_{\rm hol}(A,R),\label{EFF2}
\end{eqnarray}
where
\begin{eqnarray}
V_a^2=(V_a^1)^2+(V_a^2)^2,
\qquad
W_a^2=(W_a^1)^2+(W_a^2)^2,
\end{eqnarray}
and
\begin{eqnarray}
V_a\cdot W_a=V_a^1W_a^1+V_a^2W_a^2,\qquad V_a\times W_a =V_a^1W_a^2-V_a^2W_a^1.
\end{eqnarray}
\medskip
\noindent
The planar collective variables are
\begin{eqnarray}
A
=\lambda\sum_{a=1}^{d}V_a\cdot W_a,
\qquad
B=\lambda
\sum_{a=1}^{d}V_a\times W_a,
\qquad
R=\sqrt{A^2+B^2},
\end{eqnarray}
where
\begin{eqnarray}
\lambda=\frac{2N}{a},
\end{eqnarray}
and \(a\) is the lattice spacing.

\medskip
\noindent
The holonomy-induced potential is then given by 

\begin{eqnarray}
V_{\rm hol}(A,R)
=
-\log\Big(I_0(R)-\frac{A}{R}I_1(R)\Big).
\label{EFF1}
\end{eqnarray}

\medskip
\noindent In the continuum limit
\begin{eqnarray}
a\to 0,
\qquad
\Lambda\to\infty,
\qquad
\beta=a\Lambda
\ \ \text{fixed},
\end{eqnarray}
the endpoint coefficients \(\alpha_\Lambda\) and \(\beta_\Lambda\), which define the mass scale
\begin{eqnarray}
m=\frac{2N}{a}\alpha_\Lambda
\end{eqnarray}
and the anisotropic endpoint coupling, are given explicitly in terms of the oscillator mass \(s\), the lattice spacing \(a\), and the fugacity \(x=e^{-\beta s}\) by
\begin{eqnarray}
\alpha_\Lambda
=
\frac12+\frac{as}{2}\frac{1+x^2}{1-x^2}+O(a^2),
\qquad 
\beta_\Lambda
=
as\,\frac{x}{1-x^2}+O(a^2),
\qquad
x=e^{-\beta s}.
\label{alphabeta_x_intro}
\end{eqnarray}

\medskip
\noindent
The physical domain is constrained by \(R^2\ge A^2\). The relevant saddle of the holonomy potential is therefore not an unconstrained saddle in the ambient \((A,R)\)-space, but a constrained boundary saddle on the aligned branch
\begin{eqnarray}
B=0,
\qquad
A=R.
\end{eqnarray}
Restricting the holonomy potential to this branch gives
\begin{eqnarray}
V_{\rm b}(A)
=
-\log\Big(I_0(A)-I_1(A)\Big).
\end{eqnarray}
The constrained saddle \(A_*=R_*\) is determined by
\begin{eqnarray}
I_0(R_*)
=
\left(1+\frac{1}{R_*}\right)I_1(R_*),
\qquad
R_*\simeq 1.545.
\end{eqnarray}
We denote
\begin{eqnarray}
V_*
=
-\log\Big(I_0(R_*)-I_1(R_*)\Big).
\end{eqnarray}

\medskip
\noindent
Expanding the holonomy potential around the constrained boundary saddle \(A=R_*\), \(B=0\), and organizing the result in terms of the longitudinal displacement \(A^2-R_*^2\) and the transverse variable \(B^2\), one obtains
\begin{eqnarray}
V_{\rm hol}(A,B)
&=&
V(A)
+
c_2(A)\,B^2
+
c_4\,B^4
+
O_3\!\left(A^2-R_*^2,\;B^2\right),
\label{Vhol_Vc2c4_AB1}
\end{eqnarray}
where
\begin{eqnarray}
V(A)
=
V_*
-\frac{R_*-1}{8R_*^3}
\Big(A^2-R_*^2\Big)^2,
\label{VA}
\end{eqnarray}
and
\begin{eqnarray}
c_2(A)
=
-\frac{1}{2R_*}
+
\frac{3-2R_*}{8R_*^3}
\Big(A^2-R_*^2\Big),
\qquad
c_4
=
\frac{3-R_*}{8R_*^3}.
\end{eqnarray}
Thus \(A^2\) parametrizes motion along the aligned branch, while \(B^2\) measures the departure from it.

\medskip
\noindent
A particularly useful transverse probe is obtained by freezing the longitudinal variable at its saddle value,
\begin{eqnarray}
A=R_*.
\end{eqnarray}
Substituting this into \eqref{Vhol_Vc2c4_AB1}, one finds
\begin{eqnarray}
V_{\rm hol}(A=R_*,B)
&=&
V_*
-\frac{1}{2R_*}\,B^2
+
\frac{3-R_*}{8R_*^3}\,B^4
+
O(B^6).
\label{Vhol_fixedAstar_B}
\end{eqnarray}
Hence, at fixed \(A=R_*\), the holonomy potential takes a Landau-type form in the transverse variable \(B\): the negative quadratic term shows that the potential decreases as one moves away from the aligned branch, while the positive quartic term gives the leading local stabilizing correction.

\subsection{The anisotropic coupling}

\medskip
\noindent
In the full endpoint theory, the anisotropic coupling \(\beta_\Lambda\) enters in two distinct ways:
\begin{eqnarray}
D_\Lambda=\alpha_\Lambda^2-\beta_\Lambda^2,
\qquad
\text{and}
\qquad
-2\beta_\Lambda A.
\end{eqnarray}
The first contribution is even and enters through the determinant--like \(D\)-channel, whereas the second is odd and couples directly to the aligned invariant \(A\).

\medskip
\noindent
In the pure \(B\)-theory (and likewise in the \(B\)-toy model), the aligned channel is frozen and the effective theory depends only on the transverse invariant \(B\).  As a result, the odd term $-2\beta_\Lambda A$  no longer survives as an independent dynamical coupling, and $\beta_\Lambda$ appears only through the combination $D_\Lambda$. The partition function therefore reduces to

\begin{eqnarray}
Z_\perp=Z_\perp(D_\Lambda),
\end{eqnarray}
with no separate dependence on \(\beta_\Lambda\).

\medskip
\noindent
In this case, the continuum limit is independent of \(\beta_\Lambda\). Indeed, the singular behavior of the \(D\)-channel is governed solely by the approach to the critical value
\begin{eqnarray}
D_\Lambda\to \frac14,
\end{eqnarray}
and, in the pure \(B\)-theory, this is already fully captured by
\begin{eqnarray}
D_\Lambda=\alpha_\Lambda^2.
\end{eqnarray}
Thus, for the purposes of the continuum singular structure, one may effectively set
\begin{eqnarray}
\beta_\Lambda=0.
\end{eqnarray}

\medskip
\noindent
A more precise view is that, in the pure \(B\)-theory, the term $-2\beta_\Lambda A$ should be included in the
full holonomy potential, and the saddle should then be determined from the full theory:
\begin{eqnarray}
V'_{\rm hol}(A,R)
=-2\beta_\Lambda\,A+V_{\rm hol}(A,R).
\end{eqnarray}
\medskip
\noindent
The constrained boundary saddle becomes \(\beta_\Lambda\)-dependent:
\begin{eqnarray}
(A_*,B_*)=(R_*,0),
\qquad
R_*=R_*(\beta_\Lambda).
\end{eqnarray}
In other words,  the aligned coupling shifts the location of the background saddle. In this sense, the coupling \(-2\beta_\Lambda A\) is irrelevant in the pure \(B\)-theory.

\medskip
\noindent
By contrast, in the pure \(A\)-theory one keeps \(A\) dynamical and expands around
\begin{eqnarray}
B_*=0.
\end{eqnarray}
Then
\begin{eqnarray}
-2\beta_\Lambda A
=
-2\beta_\Lambda(A-R_*)
-2\beta_\Lambda R_*,
\end{eqnarray}
and the first term survives as a genuine linear coupling to the fluctuation \(A-R_*\). This is why the \(\beta_\Lambda\)-channel remains physically relevant in the \(A\)-theory, but not in the pure \(B\)-theory.

\subsection{Large--\(R\) regime and aligned dominance}

\medskip
\noindent
In the low--temperature regime the invariants \(A\), \(B\), and
\(R^2=A^2+B^2\) are driven into the large-\(R\) region, as discussed in~\cite{Ydri:2026rke}. In this regime, the holonomy potential on the constrained boundary is dominated by the linear contribution \(-A\).  This follows directly from the boundary potential
\begin{eqnarray}
V_{\rm b}(R)
=
-\log\!\Big(I_0(R)-I_1(R)\Big).
\end{eqnarray}
Using the large-\(R\) asymptotics
\begin{eqnarray}
I_0(R)
&\sim&
\frac{e^R}{\sqrt{2\pi R}}
\left(
1+\frac{1}{8R}+\cdots
\right),
\\
I_1(R)
&\sim&
\frac{e^R}{\sqrt{2\pi R}}
\left(
1-\frac{3}{8R}+\cdots
\right),
\end{eqnarray}
one obtains
\begin{eqnarray}
I_0(R)-I_1(R)
\sim
\frac{e^R}{\sqrt{2\pi R}}
\left(
\frac{1}{2R}+\cdots
\right),
\end{eqnarray}
and hence
\begin{eqnarray}
V_{\rm b}(R)
=
-R+O(\log R).
\end{eqnarray}
Since \(A=R\) on the aligned boundary, this is precisely the large-\(R\) statement
\begin{eqnarray}
V_{\rm b}(R)\sim -A,
\end{eqnarray}
up to logarithmic and inverse-power corrections.

\medskip
\noindent
This shows that the Gaussian approximation should not be constructed solely from the quadratic terms, but must also incorporate this linear aligned contribution. In other words, the natural expansion point for the continuum limit is not the bare Gaussian theory, but a \emph{shifted Gaussian} in which the linear term is absorbed into the quadratic part.

\medskip
\noindent
To implement this, we add and subtract the linear term \(A\), and write
\begin{eqnarray}
S_{N=2,\rm eff}^{\rm BFSS_{d+1}}
&=&
\Big[
m\sum_{a=1}^{d}\Big((V_a)^2+(W_a)^2\Big)
-A
\Big]
+\widetilde{V}'_{\rm hol}(A,R),\label{trueGaussian}
\end{eqnarray}
where the modified holonomy potential is defined by
\begin{eqnarray}
\widetilde{V}'_{\rm hol}(A,R)
:=
A+V'_{\rm hol}(A,R).\label{trueGaussian1}
\end{eqnarray}
Thus the Gaussian part is effectively shifted by the linear aligned contribution, while the remaining holonomy potential is regularized accordingly.

\medskip
\noindent
For simplicity, we set \(\beta_\Lambda=0\).  The compensating term \(+A\) in \(\widetilde V_{\rm hol}\) should not be expanded
as an independent fluctuation around \(A=R_*\). If one then imposes
\(A=R_*\), its transverse effect is lost, although the corresponding \(-A\)
has already been absorbed into the Gaussian sector. Instead, \(+A\) must be restricted to the transverse shell
\begin{eqnarray}
R^2=A^2+B^2,\qquad R=R_*,
\end{eqnarray}
so that
\begin{eqnarray}
A=\sqrt{R_*^2-B^2}
=
R_*-\frac{B^2}{2R_*}-\frac{B^4}{8R_*^3}+O(B^6).
\end{eqnarray}
Thus
\begin{eqnarray}
\widetilde V_{\rm hol}(B)
=
V_{\rm hol}(A=R_*,B)+\sqrt{R_*^2-B^2}.
\end{eqnarray}
Using \eqref{Vhol_fixedAstar_B}, we obtain
\begin{eqnarray}
\widetilde V_{\rm hol}(B)
=
V_*+R_*
-\frac{1}{R_*}B^2
+\frac{2-R_*}{8R_*^3}B^4
+O(B^6).\label{Vhol_fixedAstar_B1}
\end{eqnarray}
Hence the addition of \(+A\) does not merely shift the potential by a constant.
It changes the transverse Landau coefficients. This is precisely the effect that
would be missed by first replacing \(A\) by \(R_*\).

\medskip
\noindent
The introduction of the linear term \(-A\) is not an arbitrary modification, but reflects the intrinsic large--\(R\) behavior of the holonomy potential. It can be viewed as a finite renormalization of the Gaussian sector, which becomes essential in order to reproduce the correct continuum scaling. In particular, this shift will later translate into an effective renormalization of the mass parameter governing the Wishart sector.

\subsection{A toy model}\label{section3}

\medskip
\noindent
The failure of finite polynomial treatments to reproduce the
required continuum behavior~\cite{Ydri:2026rke} motivates a
non-polynomial toy model that reproduces the qualitative structure
of the local transverse expansion near the saddle, yields a
nontrivial continuum limit, and captures the global shape of
the exact potential. A simple candidate is
\begin{eqnarray}
V_{\rm toy}(B):=-\log\cosh B,
\qquad
e^{-V_{\rm toy}(B)}=\cosh B.
\label{Vtoy_def_synthesis}
\end{eqnarray}
Its local expansion is
\begin{eqnarray}
V_{\rm toy}(B)
=
\sum_{n\ge 1}(-1)^n c_{2n}B^{2n}
=
-\frac12 B^2+\frac1{12}B^4-\frac1{45}B^6
+\frac{17}{2520}B^8+O(B^{10}).
\label{Vtoy_potential_series}
\end{eqnarray}
The quadratic and quartic terms have the same signs as the
local transverse holonomy expansion. Moreover,
\(V_{\rm toy}'(B)=-\tanh B\), so the potential decreases
monotonically for \(B>0\), whereas its quartic truncation
eventually turns upward. The toy model therefore provides a
qualitative non-polynomial completion, rather than an exact
reconstruction of the holonomy potential. Detailed comparisons
are given in Section~3 of the first arXiv
version~\cite{YdriGramV1}.

\medskip
\noindent
For the later Gram formulation, it is useful to organize the
interaction through a quadratic source. With
\(\langle\cdot\rangle_0\) denoting the normalized endpoint
Gaussian average, define
\begin{eqnarray}
Z_0(c):=\Big\langle e^{c B^2}\Big\rangle_0.
\label{toy_operator_rep0}
\end{eqnarray}
Powers of \(B^2\) are generated by source derivatives,
\begin{eqnarray}
\Big\langle B^{2n}e^{c B^2}\Big\rangle_0
=
\left(\frac{\partial}{\partial c}\right)^n Z_0(c).
\end{eqnarray}
The corresponding formal differential-operator representation is
\begin{eqnarray}
G_{\rm toy}
=
\left.
\exp\!\left(
\sum_{n>1}(-1)^{n+1}c_{2n}
\left(\frac{\partial}{\partial c}\right)^n
\right)
Z_0(c)
\right|_{c=c_2}.
\label{toy_operator_rep}
\end{eqnarray}
This representation generates the higher-order interactions
from the quadratic kernel and will be used in the Gram
formulation below.

\medskip
\noindent
The full toy interaction can instead be averaged directly.
Expanding its Boltzmann weight,
\begin{eqnarray}
\cosh B
=
\sum_{n=0}^\infty \frac{B^{2n}}{(2n)!},
\label{Vtoy_full_series0}
\end{eqnarray}
and using the Gaussian moments
\begin{eqnarray}
\langle B^{2n}\rangle_0
=
\frac{(2n)!\,(d)_n}{4^n n!\,D_\Lambda^n},
\end{eqnarray}
one obtains
\begin{eqnarray}
G_{\rm toy}(D_\Lambda)
=
\langle\cosh B\rangle_0
=
\sum_{n=0}^\infty \frac{(d)_n}{n!}
\left(\frac{1}{4D_\Lambda}\right)^n
=
\left(1-\frac{1}{4D_\Lambda}\right)^{-d}.
\end{eqnarray}
Here \((d)_n=d(d+1)\cdots(d+n-1)\), for \(n\ge1\), is the rising factorial and
\(D_\Lambda=\alpha_\Lambda^2-\beta_\Lambda^2\).
The series converges for \(D_\Lambda>\tfrac14\), and its sum
becomes singular as \(D_\Lambda\to\tfrac14^+\).

\medskip
\noindent
To identify the thermal contribution of this singularity,
we normalize the interaction factor at \(x=0\), where
\(x=e^{-\beta s}\):
\begin{eqnarray}
\frac{G_{\rm toy}(D_\Lambda(x))}{G_{\rm toy}(D_0)}
=
1+\widehat{\mathcal C}_2^{\rm toy}x^2+O(x^4),
\qquad
D_\Lambda(x)=D_0+D_2x^2+O(x^4).
\end{eqnarray}
Thus \(\widehat{\mathcal C}_2^{\rm toy}\) is the coefficient
of \(x^2\) in the normalized interaction factor, with the
Gaussian partition function divided out. Following the
endpoint analysis~\cite{Ydri:2026rke}, it is
\begin{eqnarray}
\widehat{\mathcal C}_2^{\rm toy}
=
D_2\frac{G'_{\rm toy}(D_0)}{G_{\rm toy}(D_0)}
=
-\frac{dD_2}{4D_0^2\left(1-\frac{1}{4D_0}\right)}.
\end{eqnarray}
The singular logarithmic derivative compensates the vanishing
of \(D_2\) in the continuum limit, yielding
\begin{eqnarray}
\widehat{\mathcal C}_2^{\rm toy}
\longrightarrow -2d,
\qquad (\mu\to0).
\end{eqnarray}
The toy model therefore reproduces the continuum
\(D\)-channel coefficient. A finite truncation of the moment
series remains regular at \(D_\Lambda=\tfrac14\) and cannot
reproduce this mechanism: the singularity emerges only after
summing the infinite tower.

\medskip
\noindent
The toy model recovers only the \(D\)-channel contribution.
The exact endpoint theory also contains a \(\beta\)-channel
contribution~\cite{Ydri:2026rke},
\begin{eqnarray}
\frac{\beta_1^2}{2}\,
\frac{\partial_{\beta_\Lambda}^2G_{\rm ex}(D_0,0)}
{G_{\rm ex}(D_0,0)},
\qquad
\beta_\Lambda(x)=\beta_1x+O(x^3).
\end{eqnarray}
This contribution is absent from the toy model because
\(G_{\rm toy}\) depends on \(D_\Lambda\) alone, with no
independent dependence on \(\beta_\Lambda\).

\medskip
\noindent
The connection with the aligned \(-A\) theory follows from
the more general Gaussian identity
\begin{eqnarray}
\big\langle e^{uA}\cosh(vB)\big\rangle_0
=
\left(
\frac{D_\Lambda}
{D_\Lambda-\beta_\Lambda u-\frac{u^2+v^2}{4}}
\right)^d,
\qquad
D_\Lambda=\alpha_\Lambda^2-\beta_\Lambda^2.
\label{gaussian_master}
\end{eqnarray}
Here \(u\) and \(v\) are insertion parameters, and the identity
holds where the Gaussian integrals converge. In particular,
for \(u=-1\), \(v=1\), and \(\beta_\Lambda=\tfrac12\),
\begin{eqnarray}
\big\langle e^{-A}\cosh B\big\rangle_0=1.
\label{toy_trivialization}
\end{eqnarray}
Equivalently,
\begin{eqnarray}
\int dV\,dW\;
e^{-\alpha_\Lambda(V^2+W^2)}\,
\cosh B
=
\int dV\,dW\;
e^{-\alpha_\Lambda(V^2+W^2)+A}.
\label{remarkable}
\end{eqnarray}
This identifies the toy interaction with the aligned
\(-A\) interaction after Gaussian integration, not pointwise.
The general effective-\(A\) shift and its dependence on
\(\beta_\Lambda\) are discussed in Section~3.3
of~\cite{YdriGramV1}.

\medskip
\noindent
Once the aligned contribution \(-A\) is absorbed into the
Gaussian action, the remaining toy interaction contains the
compensating term \(+A\). For the local transverse completion,
this term is evaluated on the fixed-radius background
\(A=\sqrt{R_*^2-B^2}\), giving
\begin{eqnarray}
\boxed{
\widetilde V_{\rm toy}(B)
=
V_{\rm toy}(B)+A
=
-\log\cosh B+\sqrt{R_*^2-B^2}
}.
\label{Vtoy_modified_series20}
\end{eqnarray}
The Gaussian identity \eqref{remarkable} concerns the
unconstrained endpoint integral; the fixed-radius substitution
specifies the local completion used below. Its expansion is
\begin{eqnarray}
\widetilde V_{\rm toy}(B)&=&
R_*+\sum_{n\ge1}(-1)^n c'_{2n}B^{2n}\nonumber\\
&=&
R_*
-\left(\frac12+\frac{1}{2R_*}\right)B^2
+\left(\frac1{12}-\frac{1}{8R_*^3}\right)B^4
-\left(\frac1{45}+\frac{1}{16R_*^5}\right)B^6
\nonumber\\
&+&
\left(\frac{17}{2520}-\frac{5}{128R_*^7}\right)B^8
+O(B^{10}).
\label{Vtoy_modified_series1}
\end{eqnarray}
The modified coefficients \(c'_{2n}\) enter the later
Gram/Wishart interaction, while \eqref{remarkable} and the
singular summed Gaussian average provide the benchmarks for
its resummed treatment.

\section{Gram matrix formulation}\label{section4}

\subsection{Big Gram matrix and exact decomposition of the holonomy invariant}

\medskip
\noindent
The elementary Gram-matrix calculus underlying the following construction is detailed in the first arXiv version of this paper~\cite{YdriGramV1}. There we define the endpoint matrices \(X_0\) and \(X_1\), the cross-covariance matrix \(K=X_1X_0^T\), and the Gram blocks \(Q^V\), \(Q^W\), and \(\Phi\). We also prove the identities expressing the holonomy invariants \(A_0\), \(B_0\), and \(R_0^2=A_0^2+B_0^2\) in terms of these Gram blocks, including the decomposition into the Frobenius piece \(\mathrm{Tr}(Q^WQ^V)\) and the minor piece \((\mathrm{Tr}\,\Phi)^2-\mathrm{Tr}(\Phi^2)\). In the main text we take these identities as input and proceed directly to the Gram matrix formulation.

\medskip
\noindent
The Wishart/Stiefel decomposition used below is standard in multivariate statistics and invariant integration on matrix spaces. It goes back to Wishart's covariance distribution, Stiefel's frame manifolds, and the work of James and Herz on orthogonal invariance, latent roots, and matrix-variate special functions~\cite{Wishart:1928,BStiefel:1935,James:1954AMS,Herz:1955,James:1960AMS,James:1961AMS,James:1964AMS,Muirhead:1982}.

\bigskip
\noindent
\textbf{Big Gram matrix.}

\bigskip
\noindent
After integrating out the bulk fields, the remaining dynamical variables
are the two boundary vectors
\begin{eqnarray}
V_a^\mu,\qquad W_a^\mu ,
\end{eqnarray}
where $\mu=1,2$ labels the planar directions and
$a=1,\dots,d$ labels the $d$ matrix species.
Thus at each boundary point we have $d$ planar vectors.

\medskip
\noindent
It is convenient to assemble these vectors into the planar endpoint
$2\times d$ matrices,
\begin{eqnarray}
(X_0)_{\mu a}:=V_a^\mu,
\qquad
(X_1)_{\mu a}:=W_a^\mu,
\qquad
\mu=1,2,\quad a=1,\dots,d .
\end{eqnarray}
Equivalently, by $O(d)$ invariance the endpoint data may be encoded in the
$d\times d$ Gram blocks together with the cross--Gram block,
\begin{eqnarray}
Q^V:=X_0^T X_0,
\qquad
Q^W:=X_1^T X_1,
\qquad
\Phi:=X_1^T X_0 .
\label{Qblocks_def0}
\end{eqnarray}
By construction $Q^V$ and $Q^W$ are positive semidefinite (PSD) matrices, and
\begin{eqnarray}
\mathrm{rank}(Q^V),\;\mathrm{rank}(Q^W)\le 2,
\end{eqnarray}
since $X_0$ and $X_1$ have only two rows. The cross--Gram block $\Phi$ is
generally \emph{not} symmetric, as it couples two distinct endpoint sets.

\medskip
\noindent
The natural object that packages all constraints is the \emph{big Gram matrix}
\begin{eqnarray}
{\cal Q}:=
\begin{pmatrix}
Q^W & \Phi\\
\Phi^T & Q^V
\end{pmatrix}
=
\begin{pmatrix}
X_1^T\\ X_0^T
\end{pmatrix}
\begin{pmatrix}
X_1 & X_0
\end{pmatrix}
= Z^T Z,
\qquad
Z:=\big(X_1~X_0\big)\in\mathbb{R}^{2\times (2d)}.
\label{bigQ_def}
\end{eqnarray}
Thus ${\cal Q}$ is positive semidefinite with rank at most two,
\begin{eqnarray}
{\cal Q}\succeq 0,
\qquad
\mathrm{rank}({\cal Q})\le 2 .
\label{bigQ_constraints}
\end{eqnarray}
\medskip
\noindent
\medskip
\noindent
The cross--block $\Phi$ is automatically constrained by the positivity of ${\cal Q}$.
In particular, when $Q^V$ is invertible, the condition ${\cal Q}\succeq0$ is
equivalent to the positivity of the Schur complement of $Q^V$.
This yields the matrix Cauchy--Schwarz constraint
\begin{eqnarray}
Q^W-\Phi\,(Q^V)^{-1}\Phi^T \succeq 0
\qquad\Longleftrightarrow\qquad
\Phi\,(Q^V)^{-1}\Phi^T \preceq Q^W,
\label{CS_constraint0}
\end{eqnarray}
(and similarly with $V\leftrightarrow W$ when $Q^W\succ0$).
Thus $\Phi$ cannot be treated as a completely free matrix; it must satisfy
these positivity constraints.

\medskip
\noindent
Geometrically, \eqref{CS_constraint0} is the matrix analogue of the
Cauchy--Schwarz inequality. It ensures that the cross--Gram block $\Phi$
can arise from actual vector families $V_a^\mu$ and $W_a^\mu$, since the
projection of one set of vectors onto the span of the other cannot exceed
its total norm.

\bigskip
\noindent
\textbf{Exact decomposition of the holonomy invariant.}

\bigskip
\noindent With $K=X_1 X_0^T$ we have
\begin{eqnarray}
A_0=\mathrm{tr}\,K=\mathrm{Tr}\,\Phi,\qquad
R_0^2=A_0^2+B_0^2=\mathrm{tr}(K K^T)+2\det K,
\end{eqnarray}
and the two pieces can be written purely in terms of the Gram blocks \eqref{Qblocks_def0}:
\begin{eqnarray}
\mathrm{tr}(K K^T)
&=&
\sum_{a,b=1}^d Q^W_{ab}\,Q^V_{ab}
\;=\;
\mathrm{Tr}\big(Q^W Q^V\big)\qquad
(\text{trace over }d\times d\text{ indices}),
\label{frobenius_piece}
\\
2\det K
&=&
(\mathrm{Tr}\,\Phi)^2-\mathrm{Tr}(\Phi^2)
\;=\;
2\sum_{1\le a<b\le d}
\det\!\begin{pmatrix}
\Phi_{aa} & \Phi_{ab}\\
\Phi_{ba} & \Phi_{bb}
\end{pmatrix}.
\label{minor_piece0}
\end{eqnarray}
Therefore
\begin{eqnarray}
R_0^2
&=&
\mathrm{Tr}(Q^W Q^V)
+
\Big[(\mathrm{Tr}\,\Phi)^2-\mathrm{Tr}(\Phi^2)\Big]\nonumber\\
&=&
\sum_{a,b=1}^d Q^W_{ab}\,Q^V_{ab}
+\sum_{a,b=1}^d\Big(Q_{aa}^{WV}Q_{bb}^{WV}-Q_{ab}^{WV}\,Q_{ba}^{WV}\Big).
\label{R2_full_blocks}
\end{eqnarray}
And
\begin{eqnarray}
B_0^2
&=&
\sum_{a,b=1}^d Q^W_{ab}\,Q^V_{ab}
-\sum_{a,b=1}^dQ_{ab}^{WV}\,Q_{ba}^{WV}.
\label{B2_full_blocks}
\end{eqnarray}

\medskip
\noindent
This leads to a natural three--block description of the endpoint data
$(Q^V,Q^W,\Phi)$ subject to the Gram constraints
\begin{eqnarray}
{\cal Q}=
\begin{pmatrix}
Q^W & \Phi\\
\Phi^T & Q^V
\end{pmatrix}
\succeq0,
\qquad
\mathrm{rank}({\cal Q})\le2 .
\end{eqnarray}

\medskip
\noindent
The Gaussian endpoint mass term \eqref{EFF2} is already diagonal in \((Q^V,Q^W)\),
\begin{eqnarray}
S_{\rm mass}
=
m\big(\mathrm{Tr}Q^V+\mathrm{Tr}Q^W\big),
\qquad
m=\alpha_{\Lambda}\lambda,\qquad \lambda=\frac{2N}{a}.
\end{eqnarray}

\medskip
\noindent 
The full Gaussian contribution includes not only the quadratic mass term, but also the linear aligned component induced by the holonomy sector, as explained around \eqref{trueGaussian} and \eqref{trueGaussian1}. Combining the explicit anisotropic coupling \(-2\beta_{\Lambda}A\) with the large--\(R\) asymptotic contribution \(-A\), one obtains

\begin{eqnarray}
-2\beta_{\Lambda}A - A
=
-2\Big(\beta_{\Lambda}+\tfrac{1}{2}\Big)A.
\end{eqnarray}
In rescaled variables, this becomes
\begin{eqnarray}
-2\beta'_{\Lambda}A_0 - \lambda A_0
=
-2\hat{\beta}'_{\Lambda} A_0,
\qquad
\hat{\beta}'_{\Lambda}
=
\left(\beta_{\Lambda}+\tfrac{1}{2}\right)\frac{2N}{a}.
\end{eqnarray}

\medskip
\noindent
Thus the Gaussian part of the action can be written in the compact form
\begin{eqnarray}
\boxed{
S_{\rm Gauss}
=
m\big(\mathrm{Tr}Q^V+\mathrm{Tr}Q^W\big)
-2\hat{\beta}'_{\Lambda} A_0.
}
\end{eqnarray}

\medskip
\noindent
The holonomy potential, including the contribution
\(A=\sqrt{R_*^2-B^2}\), depends only on
\(B_0^2=R_0^2-A_0^2\). Its actual transverse expansion is given by
\eqref{Vhol_fixedAstar_B1}, while \eqref{Vtoy_modified_series1}
provides a non-polynomial toy completion of this expansion, namely
\begin{eqnarray}
V_{\rm hol}(B_0)
=-\kappa_2 B_0^2
+\kappa_4 B_0^4
-\kappa_6 B_0^6
+\kappa_8 B_0^8
+\cdots,\label{Vtoy_modified_series2}
\end{eqnarray}
with
\begin{eqnarray}
\kappa_{2n}
=
c'_{2n}\bigg(\frac{2N}{a}\bigg)^n.\label{kappavscprime}
\end{eqnarray}

\medskip
\noindent
Substituting \eqref{B2_full_blocks} generates a controlled interaction between the Frobenius sector \(\mathrm{Tr}(Q^WQ^V)\) and the minor sector \eqref{minor_piece0}.

\subsection{Rank--2 Wishart/Stiefel decomposition}

The planar endpoint fields are \(2\times d\) matrices, and hence admit a decomposition under a left \(O(2)\) rotation into a diagonal \(2\times2\) part and a residual \(2\times d\) orthonormal frame in \(\mathbb{R}^d\).

\bigskip
\noindent Hence, we write the singular value decompositions
\begin{eqnarray}
X_0
=
U_0\,\mathrm{diag}(\sqrt{v_1},\sqrt{v_2})\,R_V^{T},
\qquad
X_1
=
U_1\,\mathrm{diag}(\sqrt{w_1},\sqrt{w_2})\,R_W^{T},
\label{SVD_X0X1}
\end{eqnarray}
where
\begin{eqnarray}
U_0,U_1\in O(2),
\qquad
R_V,R_W\in \mathrm{St}(2,d)\subset\mathbb{R}^{d\times 2},
\qquad
v_i,w_i\ge 0.
\end{eqnarray}
Here, $R_V$ and $R_W$ are $d\times 2$ whereas $R_V^T$ and $R_W^T$ are $2\times d$, and $\mathrm{St}(2,d)$ denotes the Stiefel manifold of orthonormal $2$--frames in $\mathbb{R}^d$: $R^T R=\mathbf{1}_2$. More explicitly, the matrices $R_V$ and $R_W$ each consist of two orthonormal vectors in $\mathbb{R}^d$:

\begin{eqnarray}
R_V=
\begin{pmatrix}
| & |\\
r_1 & r_2\\
| & |
\end{pmatrix},
\quad
r_i\cdot r_j=\delta_{ij},\qquad R_W=
\begin{pmatrix}
| & |\\
s_1 & s_2\\
| & |
\end{pmatrix},
\quad
s_i\cdot s_j=\delta_{ij}.
\end{eqnarray}

\medskip
\noindent
However, any Stiefel frame can be obtained by rotating the canonical frame
\begin{eqnarray}
E=
\begin{pmatrix}
1 &0\\
0 &1\\
0 &0\\
\vdots &\vdots
\end{pmatrix}\equiv  \binom{\mathbf{1}_2}{0}
\in\mathbb{R}^{d\times2}.
\end{eqnarray}
Thus one may write
\begin{eqnarray}
R_V = O_V\,E,
\qquad
O_V\in O(d).
\end{eqnarray}
Similarly,
\begin{eqnarray}
R_W = O_W\,E,
\qquad
O_W\in O(d).
\end{eqnarray}
The only invariant angular information is therefore the \emph{relative}
orthogonal matrix
\begin{eqnarray}
O = O_W O_V^{T}\in O(d).
\end{eqnarray}

\medskip
\noindent
Geometrically, each Stiefel matrix specifies a $2$--dimensional subspace of
$\mathbb{R}^d$:
\begin{eqnarray}
\Pi_V = \mathrm{span}(r_1,r_2),
\qquad
\Pi_W = \mathrm{span}(s_1,s_2).\label{endpoint_two_planes}
\end{eqnarray}
Thus the angular sector of the problem describes the relative orientation
of two $2$--planes inside $\mathbb{R}^d$, and the $2\times 2$ block

\begin{eqnarray}
\Omega:=\Big[O_{ij}\Big]_{i,j=1,2}\in\mathbb{R}^{2\times2}
\label{Omega_def0}
\end{eqnarray}
is precisely the $2\times2$ overlap matrix between these planes.

\medskip
\noindent
Explicitly, we have in this parameterization 
\begin{eqnarray}
Q^V=X_0^T X_0 = R_V\,\mathrm{diag}(v_1,v_2)\,R_V^{T},
\qquad
Q^W=X_1^T X_1 = R_W\,\mathrm{diag}(w_1,w_2)\,R_W^{T},
\label{QVQW_rank2}
\end{eqnarray}
and the cross--block becomes
\begin{eqnarray}
\Phi
=
X_1^{T}X_0
=
R_W\,\mathrm{diag}(\sqrt{w_1},\sqrt{w_2})\,S\,
\mathrm{diag}(\sqrt{v_1},\sqrt{v_2})\,R_V^{T},
\qquad
S:=U_1^{T}U_0\in O(2).\nonumber\\
\label{Phi_rank2}
\end{eqnarray}
\medskip
\noindent Thus only relative angular variables enter the action.
The planar $O(2)$ rotations $U_0$ and $U_1$ appear only through
their relative combination
\(
S = U_1^T U_0 \in O(2).
\)

\medskip
\noindent Moreover the action depends only on the invariants
$\mathrm{Tr}(Q^W Q^V)$, $\mathrm{Tr}\Phi$, and $\mathrm{Tr}(\Phi^2)$, and thus
it is invariant under the common left action

\begin{equation}
(R_V,R_W)\sim (O_L R_V,\;O_L R_W).
\end{equation}
After quotienting by this redundancy the only remaining angular
variable is the relative \emph{invariant} orientation
\begin{equation}
O:=O_W O_V^T\in O(d),
\end{equation}
which describes how the $2$--plane $\Pi_V$ spanned by $R_V$ is rotated
into the $2$--plane $\Pi_W$ spanned by $R_W$.

\medskip
\noindent The endpoint measure is of the Wishart/Stiefel type. For each endpoint we have the standard rank--2 Jacobian (up to an overall constant)
\begin{eqnarray}
\prod_{a=1}^d d^2 V_a
&\propto&
d\mu(R_V)\;
(v_1 v_2)^{\frac{d-3}{2}}\;|v_1-v_2|\; dv_1\,dv_2\;
d\mu(U_0),
\nonumber\\
\prod_{a=1}^d d^2 W_a
&\propto&
d\mu(R_W)\;
(w_1 w_2)^{\frac{d-3}{2}}\;|w_1-w_2|\; dw_1\,dw_2\;
d\mu(U_1).
\label{Wishart_measures_routeB0}
\end{eqnarray}
Since the action depends only on $S=U_1^T U_0$, the $U_0,U_1$ integrals reduce to a single
Haar integral over $S\in O(2)$ times an overall volume factor.

\medskip
\noindent
The holonomy term at small \(B\), with the transverse contribution
\(A=\sqrt{R_*^2-B^2}\) included, is given by
\eqref{Vtoy_modified_series2}. In the rank--2 variables this reads

\begin{eqnarray}
V_{\rm hol}(R)
&=&
-\kappa_2\,B_0^2+\kappa_4\,B_0^4+\cdots\nonumber\\
&=&
-\kappa_2\Big(\mathrm{Tr}(Q^W Q^V)-\mathrm{Tr}(\Phi^2)\Big)+\kappa_4\Big(\mathrm{Tr}(Q^W Q^V)-\mathrm{Tr}(\Phi^2)\Big)^2+\cdots.\label{Vhol_blocks_full0}
\end{eqnarray}
\medskip
\noindent Because $Q^V,Q^W$ and $\Phi$ all have rank $\le2$, the invariants entering
\eqref{Vhol_blocks_full0} depend only on the $2\times2$ overlap matrix
$\Omega$ in \eqref{Omega_def0}. For example, we compute

\begin{eqnarray}
Q^WQ^V&=&R_W\,w\,R_W^T R_V\,v\,R_V^T=R_W\,w\,\Omega\,v\,R_V^T,\qquad \Omega=E^TOE\nonumber\\
\Rightarrow\mathrm{Tr}(Q^WQ^V)&=&\mathrm{Tr}\!\Big(w\,\Omega\,v\,\Omega^T\Big)=\sum_{i,j=1}^2
w_i v_j\,(\Omega_{ij})^2.
\end{eqnarray}
Hence,  one obtains the rank--2 identity 
\begin{eqnarray}
\mathrm{Tr}(Q^W Q^V)
&=&
\sum_{i,j=1}^2 w_i v_j\,(\Omega_{ij})^2.
\label{TrQWQV_rank2}
\end{eqnarray}
Similarly, for the cross block

\begin{eqnarray}
\mathrm{Tr}\,\Phi
&=&
\mathrm{tr}\Big(\sqrt{w}\,S\,\sqrt{v}\,\Omega^{T}\Big),
\label{TrPhi_rank20}
\\
\mathrm{Tr}(\Phi^2)
&=&
\mathrm{tr}\Big[
\big(\sqrt{w}\,S\,\sqrt{v}\,\Omega^{T}\big)
\big(\sqrt{w}\,S\,\sqrt{v}\,\Omega^{T}\big)
\Big].
\label{TrPhi2_rank20}
\end{eqnarray}
Here, $\sqrt{w}:=\mathrm{diag}(\sqrt{w_1},\sqrt{w_2})$ and similarly for $\sqrt{v}$, and
$\mathrm{tr}$ denotes the $2\times2$ trace.  In particular, the minor sector becomes
\begin{eqnarray}
(\mathrm{Tr}\Phi)^2-\mathrm{Tr}(\Phi^2)
=
(\mathrm{tr}\,\phi)^2-\mathrm{tr}(\phi^2)
=
2\det\phi,
\qquad
\phi:=\sqrt{w}\,S\,\sqrt{v}\,\Omega^{T}.
\label{minor_as_detphi0}
\end{eqnarray}
This shows that the entire minor sector depends only on the oriented
area of the effective $2\times2$ overlap matrix $\phi$. This in fact is also the key simplification: \emph{the entire minor sector reduces to a single
$2\times2$ determinant built from $(S,\Omega)$ and the eigenvalues $(v_i,w_i)$.} 

\medskip
\noindent
Thus, \emph{in rank--2 variables}, we have 
\begin{eqnarray}
B_0^2
&=&\underbrace{\sum_{i,j=1}^2 w_i v_j\,(\Omega_{ij})^2}_{\mathrm{Tr}(Q^W Q^V)}
-\underbrace{\mathrm{tr}\Big(\sqrt{w}\,S\,\sqrt{v}\,\Omega^{T}\Big)^2}_{\tr\phi^2}.
\label{R2_rank2_routeB}
\end{eqnarray}

\medskip
\noindent 
\noindent
Finally, the classical endpoint Gaussian action is particularly simple in the rank--2 variables:
\begin{eqnarray}
S_{\rm mass}
=
m\big(\mathrm{Tr}Q^V+\mathrm{Tr}Q^W\big)
=
m\,(v_1+v_2+w_1+w_2).
\label{Smass_routeB0}
\end{eqnarray}
\medskip
\noindent 
Moreover, using \eqref{TrPhi_rank20}, the anisotropic coupling \(-2\beta_{\Lambda}A-A=-2\hat{\beta}'_{\Lambda}A_0\) becomes
\begin{eqnarray}
-2\beta_{\Lambda}A-A=-2\hat{\beta}'_{\Lambda}
\mathrm{tr}\phi.\label{anistropic0}
\end{eqnarray}

\subsection{The weak-minor approximation}

\medskip
\noindent Combining \eqref{Wishart_measures_routeB0}, \eqref{Smass_routeB0} and \eqref{Vhol_blocks_full0} we obtain
\begin{eqnarray}
Z_\perp
&\propto&
\int dv_1\,dv_2\;
(v_1 v_2)^{\frac{d-3}{2}}|v_1-v_2|\;
e^{-m(v_1+v_2)}
\int dw_1\,dw_2\;
(w_1 w_2)^{\frac{d-3}{2}}|w_1-w_2|\;
e^{-m(w_1+w_2)}
\nonumber\\
&&\times
\int_{O(d)} d\mu(O)\;
\int_{O(2)} d\mu(S)\;
\exp\!\Big[
\kappa_2\,B_0^2-\kappa_4B_0^4-\cdots\Big].
\label{Zperp_routeB_master}
\end{eqnarray}
\medskip
\noindent
The anisotropic coupling \eqref{anistropic0} is frozen to its longitudinal value by imposing the constrained boundary condition \(A=R_*,\ B=0\).

\medskip
\noindent
As explained around \eqref{toy_operator_rep0} and \eqref{toy_operator_rep}, the full partition function can then be generated from the quadratic kernel:

\begin{eqnarray}
\boxed{
Z_\perp
=
\left.
\exp\left(
\kappa_2\,\frac{\partial}{\partial \kappa}
-\kappa_4\frac{\partial^2}{\partial\kappa^2}
+\cdots
\right)
Z_\perp^{(2)}(\kappa)
\right|_{\kappa=0}.
}\label{ruleF}
\end{eqnarray}
Here
\begin{eqnarray}
Z_\perp^{(2)}(\kappa)&\propto&
\int dv_1\,dv_2\;
(v_1 v_2)^{\frac{d-3}{2}}|v_1-v_2|\;
e^{-m(v_1+v_2)}
\int dw_1\,dw_2\;
(w_1 w_2)^{\frac{d-3}{2}}|w_1-w_2|\;
e^{-m(w_1+w_2)}
\nonumber\\
&&\times
\int_{O(d)} d\mu(O)\;
\int_{O(2)} d\mu(S)\;
\exp\!\Big[
\kappa\,B_0^2\Big].
\label{Zperp_routeB_master_quadratic}
\end{eqnarray}
\medskip
\noindent
At fixed longitudinal background \(A=\lambda\,\mathrm{tr}\phi=R_*\), the identity \eqref{minor_as_detphi0} rewrites the invariant \eqref{R2_rank2_routeB} as

\begin{eqnarray}
\boxed{B_0^2=\sum_{i,j=1}^2 w_i v_j\,(\Omega_{ij})^2+2\det\!\Big(\sqrt{w}\,S\,\sqrt{v}\,\Omega^{T}\Big)-\frac{R_*^2}{\lambda^2}.}\label{R2_rank2_routeB0}
\end{eqnarray}

\medskip
\noindent
We now introduce the \emph{weak--minor approximation}, in which only the first term in \eqref{R2_rank2_routeB0} is retained. The \(O(2)\) variable \(S\) then drops out, and the remaining \(O(d)\) angular integral reduces to the standard \emph{orthogonal HCIZ} (Harish-Chandra--Itzykson--Zuber) generating function \cite{Harish-Chandra:1957dhy,Itzykson:1979fi,Hikami:2006wya,Brezin:2002mathph,Duistermaat:1982vw,Guhr:2000mathph,Guhr:1997ve}.

\medskip
\noindent
The second term in \eqref{R2_rank2_routeB0}, which we shall treat separately, is the genuine \emph{minor contribution}: it activates an additional \(O(2)\) integral and couples it nontrivially to the same rank--2 overlap block \(\Omega\).

\medskip
\noindent
In this approximation, the quadratic partition function reduces therefore to

\begin{eqnarray}
Z_\perp^{(2)}(\kappa)&\propto& 
\int dv_1\,dv_2\;
(v_1 v_2)^{\frac{d-3}{2}}|v_1-v_2|\;
e^{-m(v_1+v_2)}
\int dw_1\,dw_2\;
(w_1 w_2)^{\frac{d-3}{2}}|w_1-w_2|\;
e^{-m(w_1+w_2)}
\nonumber\\
&\times & e^{-\kappa\frac{R_*^2}{\lambda^2}}
\int_{O(d)} d\mu(O)\;\exp\!\Big[\kappa{\cal T}\Big],
\label{Zperp_routeB_master_quadratic0}
\end{eqnarray}
with

\begin{eqnarray}
  {\cal T}:=\sum_{i,j=1}^2 w_i v_j\,(\Omega_{ij})^2.
\end{eqnarray}

\medskip
\noindent
The remaining angular integral is
\begin{eqnarray}
{\cal I}_{\rm ang}(v,w)
:=
\int_{O(d)} d\mu(O)\;
\exp\!\Big[
\kappa\,{\cal T}
\Big].
\label{Iang_quadratic_def}
\end{eqnarray}
\medskip
\noindent
Its Haar-moment evaluation is given in Section 4.5 and Appendix B of the first arXiv version~\cite{YdriGramV1}. We retain only the quadratic result needed below. With
\begin{eqnarray}
S_1^V:=\mathrm{Tr}\Lambda_V,
\quad
S_2^V:=\mathrm{Tr}(\Lambda_V^2),
\qquad
S_1^W:=\mathrm{Tr}\Lambda_W,
\quad
S_2^W:=\mathrm{Tr}(\Lambda_W^2).
\end{eqnarray}
\begin{eqnarray}
\log{\cal I}_{\rm ang}
&=&
\frac{\kappa}{d}\,S_1^W S_1^V-\frac{\kappa^2}{2d^2}\,(S_1^W)^2(S_1^V)^2
\nonumber\\
&+&
\frac{\kappa^2}{2d(d-1)(d+2)}
\Big[
(d+1)(S_1^W)^2(S_1^V)^2
-2(S_1^V)^2S_2^W
-2(S_1^W)^2S_2^V
+2d\,S_2^W S_2^V
\Big]
\nonumber\\
&+&O({\cal T}^3).
\label{logIang_p2_explicit}
\end{eqnarray}
\medskip
\noindent
Higher orders can be organized through zonal polynomials, equivalently Jack polynomials at parameter \(\alpha=2\)~\cite{James:1960AMS,James:1961AMS,James:1965AMS}; the detailed calculation is deferred to~\cite{YdriGramV1}.

\subsection{Retaining the aligned term in the angular kernel}
\label{section5}

\medskip
\noindent
The preceding weak--minor approximation freezes the anisotropic coupling
\eqref{anistropic0} at the longitudinal background \(A=R_*\).
We now retain the aligned linear contribution before performing the
angular integration. As discussed around \eqref{trueGaussian} and
\eqref{trueGaussian1}, the holonomy potential contributes asymptotically
a term \(-A\). Setting \(\beta_\Lambda=0\), the retained linear term is
\begin{eqnarray}
-A
=-\lambda A_0=
-\ell\,\mathrm{tr}\phi,
\qquad
\ell:=\lambda.
\end{eqnarray}
As shown below, this term shifts the Gaussian coupling to
\(\alpha_\Lambda-\tfrac12\), allowing us to recover the correct continuum limit.

\medskip
\noindent
We therefore replace the purely quadratic kernel by a shifted kernel:
\begin{eqnarray}
\boxed{
Z_\perp
=
\left.
\exp\!\left(
\kappa_2\,\frac{\partial}{\partial \kappa}
-\kappa_4\frac{\partial^2}{\partial\kappa^2}
+\cdots
\right)
Z_\perp^{(2)}(\kappa,\ell)
\right|_{\kappa=0}.
}
\label{ruleF_regulated}
\end{eqnarray}
Here
\begin{eqnarray}
Z_\perp^{(2)}(\kappa,\ell)
&\propto&
\int dv_1\,dv_2\;
(v_1 v_2)^{\frac{d-3}{2}}|v_1-v_2|\;
e^{-m(v_1+v_2)}
\int dw_1\,dw_2\;
(w_1 w_2)^{\frac{d-3}{2}}|w_1-w_2|\;
e^{-m(w_1+w_2)}
\nonumber\\
&\times&
\int_{O(d)} d\mu(O)\;
\int_{O(2)} d\mu(S)\;
\exp\!\Big[\ell\mathrm{tr}\phi+\kappa{\cal T}
+2\kappa \det\!\phi-\kappa\frac{R_*^2}{\lambda^2}\Big].
\label{Zperp_routeB_master_quadratic_regulated0}
\end{eqnarray}

\medskip
\noindent
In the weak--minor approximation, we neglect the term
\(2\kappa\det\phi\) in the exponent while retaining the aligned
term \(\ell\,\mathrm{tr}\phi\). The remaining planar angular integral is

\begin{eqnarray}
\int_{O(2)} d\mu(S)\;
\exp\!\Big[\ell\mathrm{tr}\phi\Big]
&=&
\int_{O(2)} d\mu(S)\;
\exp\!\Big[\ell\mathrm{tr}SM\Big],
\qquad M:=\sqrt{v}\,\Omega^T\sqrt{w}.
\end{eqnarray}
Splitting \(O(2)\) into its two connected components gives
\begin{eqnarray}
\int_{O(2)} d\mu(S)\;e^{\ell\,\mathrm{tr}(SM)}
=
\frac12
\Big[
I_0\!\big(\ell\,\rho_+(M)\big)
+
I_0\!\big(\ell\,\rho_-(M)\big)
\Big].
\label{O2_full_integral}
\end{eqnarray}
The two arguments are defined by
\begin{eqnarray}
\rho_+(M)^2
:=
(\mathrm{tr}M)^2+(M_{12}-M_{21})^2
=
\mathrm{tr}(MM^T)+2\det M,
\label{rho_plus_def}
\end{eqnarray}
and
\begin{eqnarray}
\rho_-(M)^2
:=
(M_{11}-M_{22})^2+(M_{12}+M_{21})^2
=
\mathrm{tr}(MM^T)-2\det M.
\label{rho_minus_def}
\end{eqnarray}
Here \(\mathrm{tr}(MM^T)={\cal T}\), while
\begin{eqnarray}
\det M
=
\sqrt{v_1v_2w_1w_2}\,\det\Omega.
\label{detM_def}
\end{eqnarray}
The component-by-component calculation is given in Section~5.1
of the first arXiv version~\cite{YdriGramV1}.
The next subsection restores the determinant contribution in the
transverse interaction and combines it with this angular result.

\subsection{Improved weak--minor approximation and emergence of alignment}

\medskip
\noindent
Retaining the aligned term allows us to perform the \(O(2)\) integral
before truncating the minor sector. Using the exact factorization
\begin{eqnarray}
\det\!\Big(\sqrt{w}\,S\,\sqrt{v}\,\Omega^T\Big)
=
\det S\;\det\Omega\;\sqrt{v_1v_2w_1w_2},
\end{eqnarray}
we introduce the scalar minor variable
\begin{eqnarray}
\delta
:=
\det\Omega\;\sqrt{v_1v_2w_1w_2}=\det M.
\end{eqnarray}
The transverse invariant then reads
\begin{eqnarray}
B_0^{2}
=
{\cal T}
+2(\det S)\,\delta
-c,
\qquad
{\cal T}:=\sum_{i,j} w_i v_j(\Omega_{ij})^2,
\qquad
c:=\frac{R_*^2}{\lambda^2}.
\label{B0_split}
\end{eqnarray}

\medskip
\noindent
The regulated angular kernel is therefore
\begin{eqnarray}
K(\kappa,\ell;v,w,\Omega)
=
\int_{O(2)} d\mu(S)\;
\exp\!\Big[
\ell\,\mathrm{tr}(SM)
+\kappa\big({\cal T}+2(\det S)\delta-c\big)
\Big].
\end{eqnarray}
Splitting \(O(2)\) into its two components gives
\begin{eqnarray}
K
=
\frac12\,e^{\kappa({\cal T}-c)}
\Big[
e^{2\kappa\delta}\,I_0\!\big(\ell\,\rho_+\big)
+
e^{-2\kappa\delta}\,I_0\!\big(\ell\,\rho_-\big)
\Big],
\label{K_full_O2}
\end{eqnarray}
where
\begin{eqnarray}
\rho_\pm^2
=
\mathrm{tr}(MM^T)\pm2\det M
=
{\cal T}\pm2\delta.
\end{eqnarray}

\medskip
\noindent
The kernel is even in \(\delta\). Keeping its zeroth-order term
defines the neutral minor truncation
\begin{eqnarray}
\delta\longrightarrow0.
\end{eqnarray}
This is a truncation of the kernel, rather than a constraint on the
angular configurations; the even minor corrections are neglected.
Since \(\rho_+=\rho_-=\sqrt{\cal T}\) at this order,
\eqref{K_full_O2} reduces to the \emph{improved weak--minor kernel}
\begin{eqnarray}
\boxed{
K
=
e^{\kappa({\cal T}-c)}\;
I_0\!\big(\ell\sqrt{\cal T}\big).
}
\label{K_improved}
\end{eqnarray}

\medskip
\noindent
The Bessel factor retains the effect of the aligned term.
For large argument,
\begin{eqnarray}
I_0\!\big(\ell \sqrt{\cal T}\big)
\sim
\frac{\exp\!\big(\ell \sqrt{\cal T}\big)}
{\sqrt{2\pi \ell \sqrt{\cal T}}}.
\end{eqnarray}
With \(\ell=\lambda\), matching its dominant exponential to
\(\exp(R_*)\), obtained by imposing \(A=R_*\) directly, requires
\begin{eqnarray}
\sqrt{{\cal T}_*}=\frac{R_*}{\lambda},
\qquad
\text{i.e.}\qquad
{\cal T}_*=\frac{R_*^2}{\lambda^2}=c,
\end{eqnarray}
which is precisely \(B_{0,*}^2=0\) in the neutral truncation.
This matching concerns the exponential contribution; the Bessel
prefactor remains part of the angular measure.

\medskip
\noindent
The emergence of alignment can also be seen before the minor
truncation. At fixed ordered eigenvalues
\(v_1\ge v_2\ge0\) and \(w_1\ge w_2\ge0\), the overlap matrix
\(\Omega\) is a contraction. Consequently,
\(|\delta|\le\sqrt{v_1v_2w_1w_2}\) and
\({\cal T}\le w_1v_1+w_2v_2\), giving
\begin{eqnarray}
\rho_+^2
=
{\cal T}+2\delta
\le
{\cal T}+2|\delta|
\le
w_1v_1+w_2v_2+2\sqrt{v_1v_2w_1w_2}
=
\big(\sqrt{w_1v_1}+\sqrt{w_2v_2}\big)^2.
\label{rho_plus_bound_final}
\end{eqnarray}
The bound is saturated by an aligned configuration,
\begin{eqnarray}
\Omega_*=\mathbf{1}_2.
\end{eqnarray}
Thus alignment maximizes the dominant Bessel exponential.
The detailed bounds and saturation conditions are given in
Section~5.3 of the first arXiv version~\cite{YdriGramV1}.
This angular maximization is distinct from the matching condition
\({\cal T}_*=c\) imposed above after the neutral truncation.

\section{The \(-A\) model in the Gram/Wishart/Stiefel formulation}\label{section6}

\subsection{The kernel \(\mathcal K_d\)}

\medskip
\noindent
We now study the pure \(-A\) model in the Gram/Wishart/Stiefel
formulation. The essential issue is how the angular kernel modifies
the apparent doubled Wishart entropy, discussed below, so that the correct continuum
coefficient is recovered. The endpoint data are assembled into the
\(2\times d\) matrices
\begin{eqnarray}
(X_0)_{\mu a}:=V^\mu_a,
\qquad
(X_1)_{\mu a}:=W^\mu_a,
\qquad
\mu=1,2,
\qquad
a=1,\dots,d,
\end{eqnarray}
with associated rank--\(2\) Gram blocks
\begin{eqnarray}
Q^V=X_0^T X_0,
\qquad
Q^W=X_1^T X_1,
\qquad
\Phi=X_1^T X_0.
\end{eqnarray}
Thus,
\begin{eqnarray}
{\rm Tr}(X_0^T X_0)=\sum_a (V_a)^2,
\qquad
{\rm Tr}(X_1^T X_1)=\sum_a (W_a)^2,
\qquad
{\rm Tr}(X_1^T X_0)=\sum_a W_a\!\cdot V_a.
\end{eqnarray}
In this parametrization, the relative angular dependence is encoded in the
\(2\times2\) principal block \(\Omega\) of the relative orthogonal matrix.

\medskip
\noindent
After performing the exact \(O(2)\) integral, the pure \(-A\) source leads to the
\(O(d)\) angular kernel
\begin{eqnarray}
\mathcal K_d(v,w;\ell)
&=&
\int_{O(d)}d\mu(O)\;\frac12\Big[
I_0\!\big(\ell\,\rho_+(\Omega)\big)
+I_0\!\big(\ell\,\rho_-(\Omega)\big)
\Big],
\label{Kd_two_branch}
\end{eqnarray}
where the two branches are defined by
\begin{eqnarray}
\rho_\pm^2(\Omega)
&=&
{\cal T}(\Omega)\pm 2\,\delta(\Omega),
\nonumber\\
{\cal T}
&=&
\sum_{i,j=1}^2 w_i v_j\,\Omega_{ij}^2,
\nonumber\\
\delta=\det\phi&=&
\sqrt{w_1w_2v_1v_2}\,\det\Omega.
\label{rho_branches_def}
\end{eqnarray}

\medskip
\noindent
The large-\(\ell\) saddle selects the aligned configuration
\begin{eqnarray}
\Omega_*=\mathbf 1_2,
\end{eqnarray}
for which the dominant exponential behavior is
\begin{eqnarray}
I_0\!\big(\ell\sqrt{{\cal T}(\Omega)}\big)
\;\sim\;
\exp\!\Big[\ell(\sqrt{w_1v_1}+\sqrt{w_2v_2})\Big].
\label{I0_large_arg_aligned}
\end{eqnarray}
As will become clear below, this exponential behavior is in fact not tied to the special case
\(\delta=0\), but persists for the full kernel: the role of \(\delta\) is to modify the prefactor
and subleading structure, not the leading aligned exponential growth itself.

\medskip
\noindent
However, this also creates a problem. Indeed, freezing \(\Omega=\mathbf 1_2\) before
performing the \(O(d)\) integral leaves intact the full Wishart/Stiefel Jacobian
\begin{eqnarray}
(w_1w_2)^{\frac{d-3}{2}}|w_1-w_2|,
\label{Wishart_block_recap}
\end{eqnarray}
and one is then led to a reduced saddle-point action containing an \(O(d)\) logarithmic entropy
of the form
\begin{eqnarray}
-\frac{d-3}{2}(\log w_1+\log w_2)-\log|w_1-w_2|,
\label{Wishart_entropy_block_recap}
\end{eqnarray}
which survives into the symmetric reduced action. This produces an apparent doubled Wishart
entropy and leads to a coefficient roughly of order \(2d\) in front of the reduced logarithm.
But the exact Cartesian evaluation, given in Section~6.2 of the first
arXiv version~\cite{YdriGramV1}, shows that the pure
\(-A\) theory contains only a single soft Gaussian block, and therefore
only a single exponent \(d\), not \(2d\).

\medskip
\noindent
This means that the exact \(O(d)\) angular integral cannot be a spectator. On the contrary, it
must itself carry a highly nontrivial prefactor whose role is precisely to cancel one whole
endpoint Wishart entropy block. This may be seen very explicitly by fixing \(X_0\) and
integrating \(X_1\) in two different ways.

\medskip
\noindent
In Cartesian variables the integral over \(X_1\) can be done immediately to find
\begin{eqnarray}
\int dX_1\;
\exp\!\Big[
-m\,{\rm Tr}(X_1^T X_1)+\lambda\,{\rm Tr}(X_1^T X_0)
\Big]
&=&
\left(\frac{\pi}{m}\right)^d
\exp\!\left[
\frac{\lambda^2}{4m}{\rm Tr}(X_0^T X_0)
\right]\nonumber\\
&=&\left(\frac{\pi}{m}\right)^d
\exp\!\left[
\frac{\lambda^2}{4m}(v_1+v_2)
\right].
\label{X1_cartesian_exact_identity}
\end{eqnarray}

\medskip
\noindent
Rewriting the same integral in Wishart/Stiefel variables gives the exact identity
\begin{eqnarray}
\boxed{
\left(\frac{\pi}{m}\right)^d
\exp\!\left[
\frac{\lambda^2}{4m}(v_1+v_2)
\right]
=
C_d
\int_0^\infty dw_1\,dw_2\;
(w_1w_2)^{\frac{d-3}{2}}|w_1-w_2|\,e^{-m(w_1+w_2)}
\,
\mathcal K_d(v,w;\ell).}\nonumber\\
\label{exact_identity_cartesian_vs_wishart}
\end{eqnarray}
Since the left-hand side contains no second Wishart entropy block, the full \(w\)-dependence
of the exact angular kernel \(\mathcal K_d(v,w;\ell)\) must supply precisely the inverse
prefactor needed to remove \eqref{Wishart_block_recap}.

\subsection{Symmetry property of the kernel \(\mathcal K_d\)}

\medskip
\noindent
As it is obvious, the full Cartesian Gaussian integral is symmetric under
\begin{eqnarray}
X_0 \longleftrightarrow X_1,
\qquad
v \longleftrightarrow w.
\end{eqnarray}
However, once one passes to the mixed Wishart/Stiefel representation, one must choose which endpoint
data are held fixed and which are integrated out. These two choices must lead to equivalent integral
representations of the same object.

\medskip
\noindent
Fixing \(v\) and integrating over \(w\), one obtains
\begin{eqnarray}
C_d
\int {\cal M}(w)\,
{\cal K}_d^{(v|w)}(v,w;\ell)
=
\left(\frac{\pi}{m}\right)^d
\exp\!\left[\frac{\lambda^2}{4m}(v_1+v_2)\right].
\label{K_v_given_w}
\end{eqnarray}
Conversely, fixing \(w\) and integrating over \(v\), one has
\begin{eqnarray}
C_d
\int {\cal M}(v)\,
{\cal K}_d^{(w|v)}(w,v;\ell)
=
\left(\frac{\pi}{m}\right)^d
\exp\!\left[\frac{\lambda^2}{4m}(w_1+w_2)\right].
\label{K_w_given_v}
\end{eqnarray}
Here
\begin{eqnarray}
{\cal M}(z)
:=
(z_1z_2)^{\frac{d-3}{2}}|z_1-z_2|\,e^{-m(z_1+z_2)},
\end{eqnarray}
and \({\cal K}_d^{(v|w)}(v,w;\ell)\) is given by equations
\eqref{Kd_two_branch} and \eqref{rho_branches_def}.

\medskip
\noindent
In fact, the two mixed kernels coincide exactly after exchange of the two endpoint sectors.

\medskip
\noindent
To make this statement precise, one must examine the pushforward of the Haar measure on \(O(d)\) to the \(2\times 2\) principal block \(\Omega\). From the Stiefel reduction, this induced measure takes the form
\begin{eqnarray}
d\nu_d(\Omega)
\;\propto\;
\big(\det(\mathbf 1_2-\Omega^T\Omega)\big)^{\frac{d-3}{2}}\,d\Omega,
\qquad
\Omega^T\Omega\le \mathbf 1_2.
\end{eqnarray}
This measure is invariant under transpose. Indeed, the domain is preserved since
\begin{eqnarray}
\Omega^T\Omega\le \mathbf 1_2
\quad\Longleftrightarrow\quad
\Omega\Omega^T\le \mathbf 1_2,
\end{eqnarray}
while the density satisfies
\begin{eqnarray}
\det(\mathbf 1_2-\Omega^T\Omega)
=
\det(\mathbf 1_2-\Omega\Omega^T),
\end{eqnarray}
and the Lebesgue measure obeys \(d\Omega=d(\Omega^T)\). Hence
\begin{eqnarray}
d\nu_d(\Omega)=d\nu_d(\Omega^T).
\end{eqnarray}
Since the kernel depends on \(\Omega\) only through
\begin{eqnarray}
\sum_{i,j=1}^2 w_i v_j\,\Omega_{ij}^2,
\qquad
\det\Omega,
\end{eqnarray}
it follows that exchanging \(v\leftrightarrow w\) is equivalent to the change of variables \(\Omega\to\Omega^T\), which leaves the integral invariant. Therefore
\begin{eqnarray}
{\cal K}_d^{(v|w)}(v,w;\ell)
=
{\cal K}_d^{(w|v)}(w,v;\ell).
\end{eqnarray}

\subsection{Ansatz for \(\mathcal K_d\)}

\medskip
\noindent
As will be shown below, the dominant configuration is the symmetric saddle
\begin{eqnarray}
w_i=v_i,
\end{eqnarray}
where the distinction between the two endpoint sectors disappears. This diagonal locus is already suggested by the symmetry of the kernel \({\cal K}_d(v,w;\ell)\) under \(v\leftrightarrow w\), and is further enforced by the aligned angular configuration \(\Omega_2=\mathbf 1_2\), which drives the two rank--\(2\) endpoint frames to coincide. In the mixed representation where one sector is held fixed, this alignment translates directly into the saddle \(w_i=v_i\).

\medskip
\noindent
These considerations lead to the symmetric global ansatz
\begin{eqnarray}
\boxed{\mathcal K_d(v,w;\ell)
\sim
\frac{
\exp\!\Big[\ell(\sqrt{w_1v_1}+\sqrt{w_2v_2})\Big]
}{
(v_1v_2)^{\frac{d-3}{4}}|v_1-v_2|^{1/2}\,
(w_1w_2)^{\frac{d-3}{4}}|w_1-w_2|^{1/2}
}
\times
\mathcal P(v,w;\ell).}
\label{Kd_structural_ansatz0}
\end{eqnarray}
This is manifestly symmetric under \(v\leftrightarrow w\), and where \(\mathcal P(v,w;\ell)\) contains only subleading contributions. This form is dictated both by symmetry and by the requirement—verified in the subsequent section—that it reproduces the correct continuum scaling of the free energy.

\medskip
\noindent
In the evaluation of \eqref{exact_identity_cartesian_vs_wishart}, the integral localizes near the symmetric saddle \(v=w\), so that the kernel is effectively probed only in a neighborhood of this locus. On the diagonal, the above symmetric ansatz reduces to
\begin{eqnarray}
\mathcal K_d(v,w;\ell)
\sim
\frac{
\exp\!\Big[\ell(\sqrt{w_1v_1}+\sqrt{w_2v_2})\Big]
}{
(w_1w_2)^{\frac{d-3}{2}}|w_1-w_2|
}
\times
\mathcal P(v,w;\ell),
\qquad
w\simeq v,
\label{Kd_structural_ansatz}
\end{eqnarray}
The exponential factor is essentially fixed by the
large-argument Bessel asymptotics together with alignment, while the effective inverse Wishart block is dictated by
\eqref{exact_identity_cartesian_vs_wishart}.

\medskip
\noindent
Substituting \eqref{Kd_structural_ansatz} into \eqref{exact_identity_cartesian_vs_wishart}, the Jacobian cancels and the leading contribution reduces to
\begin{eqnarray}
\int_0^\infty dw_1\,dw_2\;
\exp\!\Big[
-m(w_1+w_2)+\ell\sqrt{w_1v_1}+\ell\sqrt{w_2v_2}
\Big].
\end{eqnarray}
This factorizes. Writing \(w_i=t_i^2\), the saddle at fixed \(v\) is
\begin{eqnarray}
\sqrt{w_i}=\frac{\ell\sqrt{v_i}}{2m},
\end{eqnarray}
and therefore
\begin{eqnarray}
\int_0^\infty dw_i\;
\exp\!\Big[-mw_i+\ell\sqrt{w_iv_i}\Big]
\sim
{\rm const}\times
\exp\!\left[\frac{\ell^2}{4m}v_i\right].
\end{eqnarray}
Hence
\begin{eqnarray}
\mathcal I(v_1,v_2)
\sim
{\rm const}\times
\exp\!\left[\frac{\ell^2}{4m}(v_1+v_2)\right],
\end{eqnarray}
which reproduces the exact Gaussian result for \(\ell=\lambda\).

\medskip
\noindent
The conclusion is that the aligned configuration remains the correct saddle, but the full
\(O(d)\) angular integral must still be retained. Its prefactor cancels one complete Wishart entropy block,
thereby resolving the doubled-entropy problem and reproducing the correct planar coefficient \(-2d\).

\medskip
\noindent
Thus the pure \(-A\) sector appears to encode a distinguished rank--\(2\) orthogonal angular kernel:
its exponential part is fixed by the aligned saddle, while its prefactor reconstructs the inverse Wishart structure
required by \eqref{exact_identity_cartesian_vs_wishart}. This strongly suggests a nontrivial orthogonal
HCIZ/matrix-Bessel identity adapted to the planar problem.

\subsection{Effective action and saddle point for the \(-A\) model}

\medskip
\noindent
From \eqref{exact_identity_cartesian_vs_wishart}, one immediately obtains the fully integrated identity
\begin{eqnarray}
\boxed{
C_d^2\int {\cal M}(v)\int {\cal M}(w)\,
{\cal K}_d(v,w;\ell)
=
\left(
\frac{\pi^2}{\lambda^2\left(\alpha_\Lambda^2-\frac14\right)}
\right)^d
}
\label{exact_identity_cartesian_vs_wishart2}
\end{eqnarray}

\medskip
\noindent
We now show that the symmetric ansatz \eqref{Kd_structural_ansatz0} reproduces the correct leading continuum behavior.

\medskip
\noindent
Substituting \eqref{Kd_structural_ansatz0} into
\eqref{exact_identity_cartesian_vs_wishart2}, one obtains, up to subleading contributions from
\(\mathcal P\),
\begin{eqnarray}
I
&\sim&
\int dv_1\,dv_2\,dw_1\,dw_2\;
\exp\!\big[-S_{\rm eff}(v,w)\big],
\end{eqnarray}
with effective action
\begin{eqnarray}
S_{\rm eff}(v,w)
&=&
m(v_1+v_2+w_1+w_2)
-\ell(\sqrt{w_1v_1}+\sqrt{w_2v_2})
\nonumber\\
&-&
-\frac{d-3}{4}\big(\log v_1+\log v_2+\log w_1+\log w_2\big)
-\frac12\log|v_1-v_2|
-\frac12\log|w_1-w_2|\nonumber\\
&-&\log \mathcal P(v,w;\ell).
\label{Seff_double_symmetric}
\end{eqnarray}

\medskip
\noindent
The saddle equations are
\begin{eqnarray}
\frac{\partial S_{\rm eff}}{\partial v_i}
=
m-\frac{\ell}{2}\sqrt{\frac{w_i}{v_i}}
-\frac{d-3}{4v_i}
-\frac{1}{2}\frac{(-1)^{i+1}}{v_1-v_2}
-\partial_{v_i}\log\mathcal P
=
0,
\label{v_saddle_eq}
\end{eqnarray}
and
\begin{eqnarray}
\frac{\partial S_{\rm eff}}{\partial w_i}
=
m-\frac{\ell}{2}\sqrt{\frac{v_i}{w_i}}
-\frac{d-3}{4w_i}
-\frac{1}{2}\frac{(-1)^{i+1}}{w_1-w_2}
-\partial_{w_i}\log\mathcal P
=
0.
\label{w_saddle_eq}
\end{eqnarray}
By symmetry under \(v\leftrightarrow w\), these equations admit the symmetric solution
\begin{eqnarray}
w_i=v_i=:z_i.
\label{symmetric_saddle_condition}
\end{eqnarray}
On this locus, the effective action reduces to
\begin{eqnarray}
S_{\rm eff}^{\rm sym}(z)
&=&
(2m-\ell)(z_1+z_2)
-\frac{d-3}{2}\big(\log z_1+\log z_2\big)
-\log|z_1-z_2|
-\log \mathcal P_{\rm sym}(z;\ell),\nonumber\\
\label{Seff_sym_reduced}
\end{eqnarray}
where
\begin{eqnarray}
\mathcal P_{\rm sym}(z;\ell):=\mathcal P(z,z;\ell).
\end{eqnarray}

\medskip
\noindent
Thus, near the dominant symmetric saddle \eqref{symmetric_saddle_condition}, the square-root reduction of the two Wishart
blocks precisely collapses to a single effective Wishart structure. In particular, if
\(\mathcal P_{\rm sym}\) is subleading in the continuum scaling, the leading dependence is governed by
\begin{eqnarray}
I
\sim
\int dz_1\,dz_2\;
(z_1z_2)^{\frac{d-3}{2}}|z_1-z_2|\,
e^{-M(z_1+z_2)},
\qquad
M:=2m-\ell.
\label{effective_single_wishart_integral}
\end{eqnarray}
This is exactly a single rank--\(2\) Wishart integral. Its overall \(M\)-dependence is immediate from
the rescaling
\begin{eqnarray}
z_i=\frac{y_i}{M},
\end{eqnarray}
which gives
\begin{eqnarray}
I
\sim
M^{-d}
\int dy_1\,dy_2\;
(y_1y_2)^{\frac{d-3}{2}}|y_1-y_2|\,e^{-(y_1+y_2)}.
\label{single_wishart_scaling}
\end{eqnarray}
Hence
\begin{eqnarray}
I\sim {\rm const}\times M^{-d}.
\label{I_scaling_M}
\end{eqnarray}

\medskip
\noindent
For the pure \(-A\) theory one has \(\ell=\lambda\), so that
\begin{eqnarray}
M
=
2m-\ell
=
\lambda(2\alpha_\Lambda-1)
=
2\lambda\left(\alpha_\Lambda-\frac12\right).
\label{M_def_pure_minusA}
\end{eqnarray}
Using
\begin{eqnarray}
\alpha_\Lambda
=
\frac12+\frac{as}{2}\frac{1+x^2}{1-x^2}+O(a^2),
\end{eqnarray}
one finds
\begin{eqnarray}
2\alpha_\Lambda-1
=
as\,\frac{1+x^2}{1-x^2}+O(a^2)
=
as+2as\,x^2+O(ax^4,a^2).
\end{eqnarray}
Therefore
\begin{eqnarray}
\log M
=
{\rm const}+2x^2+O(x^4),
\end{eqnarray}
and from \eqref{I_scaling_M} one obtains
\begin{eqnarray}
-\log I
=
d\log M+{\rm const}
=
{\rm const}+2d\,x^2+O(x^4).
\end{eqnarray}
Equivalently,
\begin{eqnarray}
I
=
{\rm const}\times \Big[1-2d\,x^2+O(x^4)\Big].
\label{I_final_minus2d}
\end{eqnarray}

\medskip
\noindent
This reproduces the coefficient \(-2d\) obtained by exact Cartesian
Gaussian integration in Section~6.2 of the first arXiv
version~\cite{YdriGramV1}. The Wishart formulation thus recovers
the same result through the reduction to a single effective
Wishart block.

\medskip
\noindent
Thus the symmetric saddle \(w_i=v_i\) is precisely what is needed: near this saddle, the
square-root reduction of the two Vandermonde/Wishart blocks collapses to a single effective
Wishart structure, and the fully integrated identity reproduces the universal continuum coefficient
\(-2d\,x^2\).

\section{Saddle-point analysis in the {\it shifted} symmetric sector}
\label{section7}

\subsection{The initial problem: angular prefactor and doubled entropy}

\medskip
\noindent
The pure \(-A\) analysis has resolved the initial problem of the
angular prefactor and the apparent doubled Wishart entropy.
The aligned configuration determines the leading exponential
behavior, but replacing the angular integrand by its value at
alignment loses the prefactor needed to recover the single-block
Gaussian benchmark.

\medskip
\noindent
The kernel ansatz \eqref{Kd_structural_ansatz0} incorporates both
contributions and reproduces the leading continuum coefficient
of the pure \(-A\) model. We now carry this structure into the
transverse theory. This requires retaining the nonperturbative
aligned kernel while treating the residual transverse angular
interaction. Whether the latter admits a controlled polynomial
truncation is a separate question, which will lead to the next
obstruction.

\subsection{Nonperturbative Bessel resummation and the shifted Wishart kernel}

\medskip
\noindent
We extend the pure \(-A\) construction to the transverse theory
by retaining its angular kernel, including the prefactor, and
organizing the residual \(\kappa\)-dependent interaction around
the same aligned configuration. The resulting Bessel/HCIZ
factorization is an ansatz, specified below. Its aligned
exponential is determined by the Bessel branch
\begin{eqnarray}
I_0\!\big(\ell\,\rho_+(\Omega)\big),
\qquad
\ell=\lambda.
\end{eqnarray}
At large \(\ell\), maximizing its exponential
\(e^{\ell\rho_+(\Omega)}\) selects the aligned configuration
\begin{eqnarray}
\Omega_*=\mathbf{1}_2,\label{ell}
\end{eqnarray}
but evaluating this exponential at alignment does not determine
the angular prefactor, which must also be retained.

\medskip
\noindent
Using the exact two-branch kernel \eqref{K_full_O2}, the angular
integral is then reorganized as follows:
\begin{eqnarray}
&&\int_{O(d)} d\mu(O)\; \int_{O(2)} d\mu(S)\;
\exp\!\Big[\ell\,{\rm tr}\phi+2\kappa \det\phi+\kappa{\cal T}-\kappa c\Big]
\nonumber\\
&=&
\int_{O(d)} d\mu(O)\;\exp\!\big[\kappa({\cal T}-c)\big]\;
\frac12\Big[
e^{2\kappa\delta}\,I_0\!\big(\ell\,\rho_+\big)
+
e^{-2\kappa\delta}\,I_0\!\big(\ell\,\rho_-\big)
\Big]
\nonumber\\
&\simeq&
\int_{O(d)} d\mu(O)\;\exp\!\big[\kappa({\cal T}-c)\big]\;
\frac12\Big[
\,I_0\!\big(\ell\,\rho_+\big)
+
\,I_0\!\big(\ell\,\rho_-\big)
\Big]\qquad \delta\longrightarrow 0
\nonumber\\
&\simeq&
\Big[\exp\!\big(\kappa({\cal T}-c)\big)\Big]_{\Omega_2=\mathbf 1_2}
\,\mathcal K_d(v,w;\ell).\label{ctcal}
\end{eqnarray}

\medskip
\noindent
In the third line, \(\delta\) has been neglected only in the exponential factors
\(e^{\pm2\kappa\delta}\). It is still retained inside the Bessel radii
\(\rho_\pm\), where
\begin{eqnarray}
\rho_\pm^2(\Omega)
=
{\cal T}(\Omega)\pm 2\,\delta(\Omega)=\sum_{i,j=1}^2 w_i v_j\,\Omega_{ij}^2
\pm 2\sqrt{w_1w_2v_1v_2}\,\det\Omega.
\end{eqnarray}
This distinction is important: the \(\kappa\)-dependent determinant splitting is
suppressed in the prefactor, while the nonperturbative Bessel sector still
remembers the two branches through \(\rho_\pm\).

\medskip
\noindent
In \eqref{ctcal}, the \(\kappa\)-dependent HCIZ factor is evaluated at the same aligned saddle selected by the nonperturbative Bessel sector, while the latter is kept under the \(O(d)\) integral and collected into the kernel \({\cal K}_d(v,w;\ell)\). At this stage one inserts the nonperturbative ansatz \eqref{Kd_structural_ansatz0} for the kernel,
\begin{eqnarray}
\mathcal K_d(v,w;\ell)
\sim
\frac{
\exp\!\Big[\ell(\sqrt{w_1v_1}+\sqrt{w_2v_2})\Big]
}{
(v_1v_2)^{\frac{d-3}{4}}|v_1-v_2|^{1/2}\,
(w_1w_2)^{\frac{d-3}{4}}|w_1-w_2|^{1/2}
},
\end{eqnarray}
thereby resumming the aligned \(-A\) sector nonperturbatively.

\medskip
\noindent
In conclusion, the residual \(\kappa\)-dependent HCIZ factor is evaluated at the same aligned configuration \eqref{ell} selected by the nonperturbative Bessel sector:
\begin{eqnarray}
e^{\kappa({\cal T}(\Omega)-c)}
\longrightarrow
e^{\kappa({\cal T}_*-c)}.
\end{eqnarray}
This is the aligned value of the HCIZ integrand, not the full Haar HCIZ
integral. Since the same configuration \(\Omega_*=\mathbf 1_2\) also maximizes
\({\cal T}(\Omega)\), the actual HCIZ integral has a saddle expansion around
this configuration:
\begin{eqnarray}
{\cal I}_{\rm HCIZ}(\kappa;v,w)
:=
\int_{O(d)}d\mu(O)\;e^{\kappa{\cal T}(\Omega)}
\sim
{\cal P}_{\rm HCIZ}(\kappa;v,w)\,
e^{\kappa{\cal T}_*}.
\label{HCIZ_recalled}
\end{eqnarray}
Thus replacing \(e^{\kappa{\cal T}_*}\) by the full HCIZ integral amounts to
restoring the associated HCIZ fluctuation prefactor.

\medskip
\noindent
In the approximation used below, we keep the nonperturbative Bessel kernel
together with this Haar-resummed treatment of the residual transverse
\(\kappa\)-sector. This gives the factorized Bessel/HCIZ ansatz
\begin{eqnarray}
&&\int_{O(d)} d\mu(O)\; \int_{O(2)} d\mu(S)\;
\exp\!\Big[\ell\,{\rm tr}\phi+2\kappa \det\phi+\kappa{\cal T}-\kappa c\Big]
\nonumber\\
&\propto&
e^{-\kappa c}
{\cal I}_{\rm HCIZ}(\kappa;v,w)\,
\frac{
\exp\!\Big[\ell(\sqrt{w_1v_1}+\sqrt{w_2v_2})\Big]
}{
(v_1v_2)^{\frac{d-3}{4}}|v_1-v_2|^{1/2}\,
(w_1w_2)^{\frac{d-3}{4}}|w_1-w_2|^{1/2}
}.
\end{eqnarray}
This should be understood as a Bessel/HCIZ factorization ansatz: the Bessel
kernel supplies the aligned \(-A\) exponential together with the
Wishart--Stiefel prefactor, while \({\cal I}_{\rm HCIZ}\) retains the
Haar-resummed transverse \(\kappa\)-interaction.

\medskip
\noindent
In the symmetric sector
\begin{eqnarray}
w_1=v_1=:z_1,
\qquad
w_2=v_2=:z_2,
\qquad
s:=z_1+z_2,
\end{eqnarray}
the aligned configuration gives
\begin{eqnarray}
\sqrt{w_1v_1}+\sqrt{w_2v_2}=z_1+z_2=s.
\end{eqnarray}
Hence the nonperturbative \(-A\) kernel contributes the exponential factor
\begin{eqnarray}
\exp\!\Big[\ell(\sqrt{w_1v_1}+\sqrt{w_2v_2})\Big]
=
e^{\ell s}.
\end{eqnarray}
Combining this with the Gaussian weight
\begin{eqnarray}
e^{-m(v_1+v_2+w_1+w_2)}
=
e^{-2ms},
\end{eqnarray}
one obtains
\begin{eqnarray}
e^{-2ms}\,e^{\ell s}
=
e^{-2m_{\rm eff}s},
\label{Gaussian_Bessel_combined}
\end{eqnarray}
with
\begin{eqnarray}
m_{\rm eff}
=
m-\frac{\ell}{2}
=
m-\frac{\lambda}{2}
=\lambda\left(\alpha_{\Lambda}-\frac{1}{2}\right).
\label{m_eff_def}
\end{eqnarray}
Equivalently,
\begin{eqnarray}
(\alpha_\Lambda)_{\rm eff}
=
\alpha_\Lambda-\frac12.
\label{alpha_eff_def}
\end{eqnarray}

\medskip
\noindent
In addition, the prefactor of the kernel reduces to
\begin{eqnarray}
(v_1v_2)^{\frac{d-3}{4}}(w_1w_2)^{\frac{d-3}{4}}
=
(z_1z_2)^{\frac{d-3}{2}},
\qquad
|v_1-v_2|^{1/2}|w_1-w_2|^{1/2}
=
|z_1-z_2|,
\end{eqnarray}
so that the logarithmic part of the effective action retains exactly the standard
Wishart and Vandermonde forms.

\medskip
\noindent
Thus, in the symmetric sector, the nonperturbative Bessel resummation simply shifts the mass
\begin{eqnarray}
m\;\longrightarrow\; m_{\rm eff},
\end{eqnarray}
while leaving the overall Wishart structure unchanged.

\subsection{The next problem: perturbative truncation and the continuum limit}

\medskip
\noindent
The shifted kernel incorporates the aligned source and its
Wishart prefactor, but the perturbative character of the remaining
transverse interaction must still be established. We test it at
the symmetric Wishart saddle, retaining only the steps needed
to identify the obstruction. The detailed quartic action,
saddle equations, and free-energy calculation are given in
Sections~7.2--7.4 of the first arXiv version~\cite{YdriGramV1}.

\medskip
\noindent
In the symmetric sector, the effective action contains the shifted
Gaussian mass term, the single Wishart/Vandermonde contribution,
and the transverse potential. We first determine the
Gaussian/Wishart saddle, which sets the scale at which the
potential must be evaluated. Introducing the eigenvalue sum,
product, and splitting,
\begin{eqnarray}
u:=\bar z_1+\bar z_2,
\qquad
p:=\bar z_1\bar z_2,
\qquad
\Delta:=\bar z_1-\bar z_2,
\end{eqnarray}
the Gaussian/Wishart action becomes
\begin{eqnarray}
S_{\rm W}(u,p,\Delta)
=
2(\alpha_\Lambda)_{\rm eff}u
-\frac{d-3}{2}\log p
-\log|\Delta|.
\label{SW_def}
\end{eqnarray}
Subtracting and adding its two eigenvalue saddle equations gives
\begin{eqnarray}
p=\frac{d-3}{4(d-2)}\,u^2,
\qquad
\Delta^2=\frac{u^2}{d-2},
\label{Wishart_relations}
\end{eqnarray}
and
\begin{eqnarray}
u_{\rm W}=\frac{d-2}{2(\alpha_\Lambda)_{\rm eff}}.
\label{uW_final}
\end{eqnarray}
Thus a small shifted coupling produces large Wishart eigenvalues,
at which the transverse expansion must remain controlled.

\medskip
\noindent
The quadratic and quartic potential terms are proportional to
\(u^2/d\) and \(u^4/d^2\), scaling at this saddle as
\(d/(\alpha_\Lambda)_{\rm eff}^{\,2}\) and
\(d^2/(\alpha_\Lambda)_{\rm eff}^{\,4}\), respectively.
More generally, the polynomial treatment organizes the even
tower schematically as~\cite{YdriGramV1}
\begin{eqnarray}
S_{\rm eff}^{\rm sym}(u)
=
2(\alpha_\Lambda)_{\rm eff}\,u
-(d-2)\log u
+\sum_{n\ge1}\frac{a_{2n}}{d^n}\,u^{2n},
\qquad
a_{2n}=O(1).
\label{Seff_u_full}
\end{eqnarray}
This coefficient counting applies at each fixed order and does
not establish convergence of the infinite series.

\medskip
\noindent
To test perturbative control in the saddle equation, we compare
the derivative of each potential term with the Gaussian/Wishart
terms. At the unperturbed saddle,
\begin{eqnarray}
\frac{u^{2n-1}}{d^n}
\sim
\frac{d^{\,n-1}}{(\alpha_\Lambda)_{\rm eff}^{\,2n-1}}.
\label{nth_potential_scaling}
\end{eqnarray}
Requiring this to be at most of order
\((\alpha_\Lambda)_{\rm eff}\) gives
\begin{eqnarray}
d^{\,n-1}\lesssim (\alpha_\Lambda)_{\rm eff}^{\,2n}.
\end{eqnarray}
Extending this power-counting requirement to arbitrarily high
orders leads to
\begin{eqnarray}
(\alpha_\Lambda)_{\rm eff}\gtrsim d^{1/2}.
\label{alpha_eff_shifted_bound}
\end{eqnarray}
Under this scaling, \(u\lesssim d^{1/2}\), and each fixed-order
potential term contributes at most \(O(1)\), compared with the
\(O(d)\) Gaussian/Wishart sector.

\medskip
\noindent
This regime is incompatible with the continuum scaling required
by the shifted Gaussian theory. Indeed, the pure \(-A\) analysis gives
\begin{eqnarray}
\alpha_\Lambda
=
\frac12+\frac{as}{2}\frac{1+x^2}{1-x^2}+O(a^2).
\end{eqnarray}
At fixed physical parameters,
\((\alpha_\Lambda)_{\rm eff}=\alpha_\Lambda-\tfrac12\)
therefore vanishes with \(a\), rather than growing as \(d^{1/2}\).
Although \(m_{\rm eff}=\lambda(\alpha_\Lambda)_{\rm eff}\)
can remain finite because \(\lambda=2N/a\), the Wishart scale
\eqref{uW_final} becomes large, and the polynomial transverse
terms are not uniformly small.

\medskip
\noindent
Resolving the angular-prefactor problem therefore exposes a
second obstruction: termwise perturbative control of the
transverse interaction is incompatible with the desired
continuum limit. This does not establish a failure of the
summed theory. We must instead retain the non-polynomial
completion and examine its resummed contribution before
taking the continuum limit.

\section{Summed local completion}\label{section8}

\subsection{The completed local transverse potential}
\subsubsection{The completed toy model revisited}
\medskip
\noindent
The bound \eqref{alpha_eff_shifted_bound} emerged from the quartic truncation as a
condition ensuring both the existence of the Wishart saddle and the perturbativity
of the transverse potential. However, this condition is incompatible with the
continuum limit, where \(\alpha_\Lambda-\tfrac{1}{2}\sim as/2\) becomes small.
This tension indicates that the bound is not intrinsic to the theory, but rather
an artifact of the polynomial truncation of the transverse sector.

\medskip
\noindent
Indeed, the truncation treats the transverse potential as a finite polynomial in
\(B^2\), whose higher-order terms grow rapidly when evaluated on the naive
Wishart scaling \(u\sim d/(\alpha_\Lambda)_{\rm eff}\). This artificial growth
forces the bound \eqref{alpha_eff_shifted_bound}. The correct procedure is instead
to resum the local transverse expansion into its non-polynomial completion.

\medskip
\noindent
The completed local transverse potential is given by \eqref{Vtoy_modified_series20},
namely
\begin{eqnarray}
\widetilde V_{\rm toy}(B)\equiv V_{\rm comp}(B)
=
-\log\cosh B+\sqrt{R_*^2-B^2}.
\label{Vcomp_B}
\end{eqnarray}
This form makes manifest that the transverse interaction is bounded and
non-polynomial, and therefore cannot exhibit the spurious large-\(d\) growth
encountered in the truncated expansion.

\medskip
\noindent
Expanding for small \(B\), one recovers the expansion \eqref{Vtoy_modified_series1},
\begin{eqnarray}
V_{\rm comp}(B)
=
R_*
-\left(\frac12+\frac{1}{2R_*}\right)B^2
+\left(\frac1{12}-\frac{1}{8R_*^3}\right)B^4
+O(B^6),
\label{Vcomp_B_expansion}
\end{eqnarray}
which defines the corrected coefficients \(c'_{2n}\). The crucial point is that
this series is only a local expansion of the bounded function
\eqref{Vcomp_B}, and its apparent divergence at large \(d\) is therefore an
artifact of truncation rather than a property of the full theory.

\subsubsection{Resumming the transverse series}

\medskip
\noindent
For the purpose of resumming the transverse series, we first localize the HCIZ factor at the aligned block
\begin{eqnarray}
\Omega_2=\mathbf 1_2.
\end{eqnarray}
Then
\begin{eqnarray}
{\cal T}\Big|_{\Omega_2=\mathbf 1_2}
=
w_1v_1+w_2v_2.
\end{eqnarray}
In the symmetric sector,
\begin{eqnarray}
w_i=v_i=z_i,
\qquad
u:=\lambda (z_1+ z_2),
\qquad
p:= \lambda^2 z_1 z_2,
\end{eqnarray}
this becomes
\begin{eqnarray}
{\cal T}
=
 z_1^2+ z_2^2
=\frac{u^2-2p}{\lambda^2}.
\end{eqnarray}
Thus the localized quadratic kernel has the form
\begin{eqnarray}
Z_{\rm loc}^{(2)}(\kappa)
=
\exp\!\left[\kappa({\cal T}-c)\right]
=
\exp\!\left[\frac{\kappa}{\lambda^2}(u^2-2p-R_*^2)\right].
\end{eqnarray}
The generating operator acts on this single variable:
\begin{eqnarray}
X:=u^2-2p-R_*^2.\label{rat}
\end{eqnarray}
Indeed,
\begin{eqnarray}
\partial_\kappa^n e^{\kappa \frac{X}{\lambda^2}}\Big|_{\kappa=0}
=\frac{X^n}{\lambda^{2n}}.
\end{eqnarray}
Therefore
\begin{eqnarray}
\exp\!\left(
\kappa_2\partial_\kappa
-\kappa_4\partial_\kappa^2
+\kappa_6\partial_\kappa^3
-\cdots
\right)
e^{\kappa \frac{X}{\lambda^2}}\Big|_{\kappa=0}
=
\exp\!\left[
c'_2X-c'_4X^2+c'_6X^3-\cdots
\right].
\end{eqnarray}
Since the coefficients \(c'_{2n}\) are those of the completed local transverse potential, the series resums to
\begin{eqnarray}
\exp[-V_{\rm comp}(X)],
\qquad
X=u^2-2p-R_*^2.
\end{eqnarray}
Equivalently, the completed transverse potential in the localized symmetric variables is
\begin{eqnarray}
V_{\rm comp}(u,p)
=
-\log\cosh B+\sqrt{R_*^2-B^2},
\label{Vcomp_up}
\end{eqnarray}
where the transverse variable is identified as
\begin{eqnarray}
B^2
=
R_*^2-u^2+2p
=
-\,X.
\end{eqnarray}
This sign is important: the localized HCIZ variable is \eqref{rat}, whereas \(B^2\) measures the transverse deviation from the aligned radius \(R_*^2\). Thus the local transverse expansion is an expansion in \(-X\), not in \(X\) itself.

\medskip
\noindent
At this stage we can use the Wishart splitting relation not as an exact saddle
condition of the completed holonomy theory, but as a diagnostic branch inherited
from the \(-A\)-shifted Gaussian problem. This is motivated by the transmutation
identity \eqref{remarkable}, which states that the completed toy model is
equivalent to the \(-A\) theory after Gaussian averaging. Thus, if the completed
model is to reproduce the same dominant region, its leading saddle structure
should be tested first on the Wishart branch
\begin{eqnarray}
p=\frac{d-3}{4(d-2)}u^2.
\end{eqnarray}
On this branch,
\begin{eqnarray}
u^2-2p
=
A_d u^2,
\qquad
A_d:=\frac{d-1}{2(d-2)}.
\end{eqnarray}
Consequently,
\begin{eqnarray}
B_{\rm W}^2(u)=R_*^2-A_d u^2,\qquad A=\sqrt{R_*^2-B_{\rm W}^2}=\sqrt{A_d}\,u.
\end{eqnarray}
\medskip
\noindent 
Thus, on the Wishart branch, the completed potential becomes
\begin{eqnarray}
V_{\rm comp}^{\rm W}(u)
=
-\log\cosh\!\left(\sqrt{R_*^2-A_d u^2}\right)
+
\sqrt{A_d} u.
\label{Vcomp_W}
\end{eqnarray}
\medskip
\noindent 
The corresponding reduced free energy is
\begin{eqnarray}
F_{\rm comp}^{\rm W}(u)
=
2(\alpha_\Lambda)_{\rm eff}u
-(d-2)\log u
-\log\cosh\!\left(\sqrt{R_*^2-A_d u^2}\right)
+
\sqrt{A_d}u.
\label{Fcomp_W}
\end{eqnarray}

\subsubsection{Flat region of the potential}

\medskip
\noindent
It is useful to separate the \(-A\) contribution from the genuinely transverse
correction. The leading \(-A\) term is kept entirely in the Gaussian sector, so
that the continuum-sensitive shift
\begin{eqnarray}
\alpha_\Lambda
\longrightarrow
\alpha_\Lambda-\frac12
\end{eqnarray}
is preserved. We therefore define
\begin{eqnarray}
M
:=
\alpha_\Lambda-\frac12.
\end{eqnarray}
The compensating \(+A\) term is then treated as part of the completed transverse
potential. Thus the reduced free energy takes the form
\begin{eqnarray}
F_{\rm comp}^{\rm W}(u)
=
2Mu
-(d-2)\log u
-\log\cosh\!\left(\sqrt{R_*^2-A_d u^2}\right)
+\sqrt{A_d}\,u .
\label{Fcomp_W1}
\end{eqnarray}

\medskip
\noindent
The remarkable identity \eqref{remarkable}, namely that the Gaussian average of
the completed potential \eqref{Vtoy_modified_series20} with respect to the
\(-A\) theory is equal to one, suggests that the dominant Wishart region of
\eqref{Fcomp_W1} is controlled by the nearly flat region of the completed
potential. Thus one expects
\begin{eqnarray}
\log\cosh\!\left(\sqrt{R_*^2-A_d u^2}\right)
\sim
\sqrt{A_d}\,u \qquad \Rightarrow \qquad \log\cosh B_{\rm W}(u)
\sim
\sqrt{R_*^2-B_{\rm W}^2(u)}.
\label{crossover}
\end{eqnarray}

\medskip
\noindent
A natural first guess would have been that the relevant region of the completed
potential is the small-\(B\) regime, since this is precisely where the local
transverse expansion is constructed. However, the exact transmutation identity
\eqref{remarkable} suggests a different picture: the dominant Wishart region
should instead be governed by the locus where the completed potential itself is
nearly zero. On the Wishart branch, this gives the crossover condition
\eqref{crossover}.

\medskip
\noindent
Indeed, this condition identifies a crossover region in which the two pieces of
the shifted completed transverse potential nearly cancel. We have 
\begin{eqnarray}
V_{\rm comp}^{\rm W}(u)
=
-\log\cosh\!\left(\sqrt{R_*^2-A_d u^2}\right)
+
\sqrt{A_d}\,u=
-\log\cosh B
+
\sqrt{R_*^2-B^2}.
\end{eqnarray}
Since
\begin{eqnarray}
 V_{\rm comp}^{\rm W}(B=0)
=
R_*
>0,
\qquad
 V_{\rm comp}^{\rm W}(B=R_*)
=
-\log\cosh R_*
<0,
\end{eqnarray}
there exists a unique \(\widetilde{B}\in(0,R_*)\) such that
\begin{eqnarray}
V_{\rm comp}^{\rm W}(\widetilde{B})=0.
\end{eqnarray}
The corresponding Wishart value is
\begin{eqnarray}
u_0
=
\sqrt{\frac{R_*^2-\widetilde{B}^2}{A_d}}.
\label{u0w}
\end{eqnarray}

\medskip
\noindent
For the physical value \(R_*\simeq 1.545\), and at large \(d\) where
\(A_d\simeq \frac12\), one finds approximately
\begin{eqnarray}
\widetilde{B}\simeq 1.36,\qquad \widetilde{A}\simeq 0.72,
\qquad
u_0\simeq 1.02.
\end{eqnarray}
In particular,
\begin{eqnarray}
\frac{\widetilde{B}}{R_*}\simeq 0.88,
\end{eqnarray}
so the relevant cancellation region lies quite close to the upper end of the
allowed \(B\)-interval, rather than near \(B=0\).

\medskip
\noindent
This is the crucial point: although one might naively expect the dominant region
to be the small-\(B\) regime because the local expansion is organized there, the
exact completed theory instead points to an intermediate/large-\(B\) region
where the two pieces of the completed potential almost cancel. In this sense,
the exact theory is not naturally controlled by a small-\(B\) expansion, even
though after Gaussian averaging it is equivalent to the \(-A\) theory.

\subsubsection{Emergence of the Wishart saddle}

\medskip
\noindent
Keeping only the leading local contribution of the completed transverse potential
around the flat point, we have
\begin{eqnarray}
 V_{\rm comp}^{\rm W}(u)
=
K_{\rm flat}(u-u_0)
+
O\!\left((u-u_0)^2\right),
\end{eqnarray}
where
\begin{eqnarray}
K_{\rm flat}
&=&
\sqrt{A_d}
\left(
1+
\frac{\widetilde A}{\widetilde B}\tanh \widetilde B
\right)
\nonumber\\
&=&1.03.
\end{eqnarray}
Here:
\begin{eqnarray}
\log\cosh \widetilde B
=
\widetilde A=
\sqrt{R_*^2-\widetilde B^2},
\qquad
u_0=\frac{\widetilde A}{\sqrt{A_d}}.
\label{flat_point_condition}
\end{eqnarray}
Hence the flat region is indeed a large-\(B\) cancellation region, but its local
linear coefficient remains of order one.

\medskip
\noindent 
Substituting this into \eqref{Fcomp_W1}, and dropping the additive constant
\(-K_{\rm flat}u_0\), gives the local Wishart form

\begin{eqnarray}
\boxed{F_{\rm comp}^{\rm W}(u)
\simeq
2M_{\rm flat}\,u
-(d-2)\log u
+
O\!\left((u-u_0)^2\right),}
\label{Fcomp_W_flat_Wishart}
\end{eqnarray}
with the shifted effective mass
\begin{eqnarray}
  \boxed{
M_{\rm flat}
=
M+\frac12K_{\rm flat}
=
\alpha_\Lambda-\frac12
+
\frac{\sqrt{A_d}}{2}
\left(
1+
\frac{\widetilde A}{\widetilde B}\tanh \widetilde B
\right).}
\label{Mflat_def}
\end{eqnarray}
Thus, near the flat cancellation region, the completed transverse potential
renormalizes the Wishart mass by the finite amount
\begin{eqnarray}
\Delta M_{\rm flat}
=
\frac12K_{\rm flat}.
\end{eqnarray}
The corresponding local Wishart saddle is therefore
\begin{eqnarray}
u_{\rm flat}
\simeq
\frac{d-2}{2M_{\rm flat}}.
\label{u_flat_saddle}
\end{eqnarray}

\subsection{Regularization and the universal \(-2d\) law}
\subsubsection{Balanced compensating split in the flat region}

\medskip
\noindent
We would like to preserve the original \(-A\) contribution entirely unchanged inside the Gaussian
sector, so that the continuum-important shift
\begin{eqnarray}
\alpha_\Lambda
\longrightarrow
\alpha_\Lambda-\frac12
\end{eqnarray}
is left untouched. The compensating \(+A\) term is then split as
\begin{eqnarray}
+A
=
(1-w)A+wA .
\end{eqnarray}
The first piece is treated in the \(A\)-representation with the Gaussian piece \(-A\), while only the second piece
is pulled back geometrically to the fixed-radius transverse shell:
\begin{eqnarray}
wA
=
w\sqrt{R_*^2-B^2}.
\end{eqnarray}

\medskip
\noindent
Thus the \(B\)-represented
completed potential becomes
\begin{eqnarray}
V_w(B)
=
-\log\cosh B
+
w\sqrt{R_*^2-B^2}.
\end{eqnarray}
The corresponding flat point \(\widetilde B_w\) is therefore no longer the old
one; it is determined by
\begin{eqnarray}
\log\cosh \widetilde B_w
=
w\,\widetilde A_w,
\qquad
\widetilde A_w
:=
\sqrt{R_*^2-\widetilde B_w^2}.
\label{flat_point_w}
\end{eqnarray}
The associated Wishart coordinate is
\begin{eqnarray}
u_w
=
\frac{\widetilde A_w}{\sqrt{A_d}}.
\end{eqnarray}

\medskip
\noindent
The slope of the \(B\)-represented piece at this new flat point is
\begin{eqnarray}
K_w^{(B)}
=
\left.
\frac{d}{du}
\left[
-\log\cosh B_{\rm W}(u)
+
w\sqrt{A_d}\,u
\right]
\right|_{u=u_w}.
\end{eqnarray}
Using
\begin{eqnarray}
\frac{dB_{\rm W}}{du}
=
-\frac{A_d u}{B_{\rm W}},
\end{eqnarray}
one obtains
\begin{eqnarray}
K_w^{(B)}
=
w\sqrt{A_d}
+
\frac{A_d u_w}{\widetilde B_w}\tanh \widetilde B_w
=
\sqrt{A_d}
\left(
w+
\frac{\widetilde A_w}{\widetilde B_w}\tanh \widetilde B_w
\right).
\label{KwB}
\end{eqnarray}

\medskip
\noindent
The remaining \(A\)-represented piece contributes the linear coefficient
\begin{eqnarray}
(1-w)u.
\end{eqnarray}
Hence the total residual linear coefficient around the \(w\)-dependent flat
point is
\begin{eqnarray}
K_w^{\rm tot}
=
(1-w)
+
\sqrt{A_d}
\left(
w+
\frac{\widetilde A_w}{\widetilde B_w}\tanh \widetilde B_w
\right).
\label{Kw_total}
\end{eqnarray}
The balanced split is therefore determined by the two conditions
\begin{eqnarray}
\log\cosh \widetilde B_w
=
w\,\widetilde A_w,
\qquad
K_w^{\rm tot}=0.
\label{balanced_w_system}
\end{eqnarray}
Equivalently, eliminating \(w\) by using
\begin{eqnarray}
w
=
\frac{\log\cosh \widetilde B_w}{\widetilde A_w},
\end{eqnarray}
one obtains the single equation
\begin{eqnarray}
1
-
\frac{\log\cosh \widetilde B_w}{\widetilde A_w}
+
\sqrt{A_d}
\left[
\frac{\log\cosh \widetilde B_w}{\widetilde A_w}
+
\frac{\widetilde A_w}{\widetilde B_w}
\tanh \widetilde B_w
\right]
=
0,
\qquad
\widetilde A_w
=
\sqrt{R_*^2-\widetilde B_w^2}.
\label{Bw_balanced_equation}
\end{eqnarray}

\medskip
\noindent
For \(R_*\simeq 1.545\) and \(A_d\simeq 1/2\), this equation gives approximately
\begin{eqnarray}
\widetilde B_w\simeq 1.527,
\qquad
\widetilde A_w\simeq 0.235,
\qquad
u_w\simeq 0.332,
\qquad
w\simeq 3.75.
\end{eqnarray}
Thus the balanced split exists, but it pushes the flat point very close to the
large-\(B\) endpoint. The above split is a large add-subtract decomposition of the
compensating \(+A\) term, not
a convex one.

\subsubsection{Resulting Wishart problem and the universal \(-2d\) law}

\medskip
\noindent
After the balanced split, the residual linear contribution generated by the
completed transverse sector is cancelled. Thus the leading effective problem
reduces to the Wishart form
\begin{eqnarray}
F_{\rm W}(u)
=
2M\,u
-
(d-2)\log u,
\qquad
M:=\alpha_\Lambda-\frac12 .
\label{FW_balanced}
\end{eqnarray}
The saddle-point equation is
\begin{eqnarray}
\frac{dF_{\rm W}}{du}
=
2M-\frac{d-2}{u}=0,
\end{eqnarray}
and hence
\begin{eqnarray}
u_{\rm W}
=
\frac{d-2}{2M}.
\label{uW_balanced}
\end{eqnarray}
The corresponding saddle free energy is
\begin{eqnarray}
F_{\rm W}(u_{\rm W})
=
(d-2)
-
(d-2)\log\left(\frac{d-2}{2M}\right).
\label{FW_saddle_balanced}
\end{eqnarray}

\medskip
\noindent
Thus the only \(M\)-dependent part of the saddle free energy is
\begin{eqnarray}
F_{\rm W}(u_{\rm W})
=
(d-2)\log(2M)
+
\hbox{\(M\)-independent terms}.
\end{eqnarray}
Therefore the normalized saddle partition function behaves as
\begin{eqnarray}
\log\frac{Z_{\rm W}(x)}{Z_{\rm W}(0)}
=
-\big(F_{\rm W}(x)-F_{\rm W}(0)\big)
=
-(d-2)\log\frac{M(x)}{M(0)}.
\label{ZW_ratio_balanced}
\end{eqnarray}

\medskip
\noindent
In the continuum low-temperature expansion,
\begin{eqnarray}
M(x)
=
\alpha_\Lambda-\frac12
=
\frac{\mu}{2}
+
\mu x^2
+
O(x^4),
\qquad
\mu:=as .
\end{eqnarray}
Hence
\begin{eqnarray}
\frac{M(x)}{M(0)}
=
1+2x^2+O(x^4),
\end{eqnarray}
and therefore
\begin{eqnarray}
\log\frac{Z_{\rm W}(x)}{Z_{\rm W}(0)}
=
-(d-2)\log\left(1+2x^2+O(x^4)\right).
\end{eqnarray}
Expanding the logarithm gives
\begin{eqnarray}
  \boxed{
\log\frac{Z_{\rm W}(x)}{Z_{\rm W}(0)}
=
-2(d-2)x^2
+
O(x^4).}
\label{minus_2d_law_balanced}
\end{eqnarray}
This is precisely the universal \(-2d\) mechanism we were after.

\subsubsection{Disappearance of the apparent bound}

\medskip
\noindent
Comparing the Wishart saddle \eqref{uW_balanced} with the center of the flat
region \eqref{u0w} gives
\begin{eqnarray}
u_{\rm W}=u_0
\qquad \Rightarrow \qquad
\frac{d-2}{2M}
=
\frac{\widetilde A}{\sqrt{A_d}} .
\label{uW_bound_estimate}
\end{eqnarray}
Equivalently,
\begin{eqnarray}
M
=
\frac{(d-2)\sqrt{A_d}}{2\widetilde A}.
\label{M_flat_matching}
\end{eqnarray}
If \(\widetilde A\) were treated as an \(O(1)\) quantity, this would imply the
apparent lower bound
\begin{eqnarray}
M\gtrsim O(d),
\end{eqnarray}
which is essentially the bound encountered in \eqref{alpha_eff_shifted_bound}.
This is incompatible with the continuum scaling
\begin{eqnarray}
M=\alpha_\Lambda-\frac12\simeq \frac{as}{2}\ll1.
\end{eqnarray}

\medskip
\noindent
The same estimate can be rewritten as a condition on the holonomy radius. Since
\begin{eqnarray}
u_0
=
\frac{\widetilde A}{\sqrt{A_d}},
\qquad
\widetilde A=\sqrt{R_*^2-\widetilde B^2}\le R_*,
\end{eqnarray}
the matching condition \(u_{\rm W}\simeq u_0\) implies
\begin{eqnarray}
\frac{d-2}{2M}
\simeq
\frac{\widetilde A}{\sqrt{A_d}}
\le
\frac{R_*}{\sqrt{A_d}},
\end{eqnarray}
and therefore
\begin{eqnarray}
R_*
\gtrsim
\frac{\sqrt{A_d}(d-2)}{2M}.
\label{Rstar_bound_from_Wishart}
\end{eqnarray}
Using \(M\simeq as/2\), this becomes, at large \(d\),
\begin{eqnarray}
R_*
\gtrsim
\frac{d}{\sqrt2\,as}.
\label{Rstar_bound_large_d}
\end{eqnarray}

\medskip
\noindent
This is not a new obstruction. It is precisely the large-\(R\) scaling already
found in the planar sector \cite{Ydri:2026rke}:
\begin{eqnarray}
R_{\rm typ}
\sim
\frac{d}{as}.
\end{eqnarray}
Thus \eqref{Rstar_bound_large_d} simply says that the flat-region radius must be
of the same order as the planar holonomy radius:
\begin{eqnarray}
R_*
\sim
R_{\rm typ}
\sim
\frac{d}{as}.
\end{eqnarray}
Equivalently, the flat-region coordinate itself scales as
\begin{eqnarray}
\widetilde A
\sim
\frac{d}{as}.
\end{eqnarray}
Substituting this scaling into \eqref{M_flat_matching} gives
\begin{eqnarray}
M
\sim
\frac{d}{d/(as)}
\sim
as,
\end{eqnarray}
which is precisely the continuum behavior.

\medskip
\noindent
It is important here to distinguish the holonomy-scaled variables from the
underlying endpoint bilinears. The large-\(R\) condition is a statement about the
variables entering the holonomy asymptotics:
\begin{eqnarray}
A=\frac{2N}{a}\widehat A,
\qquad
B=\frac{2N}{a}\widehat B,
\qquad
R=\frac{2N}{a}\widehat R .
\end{eqnarray}
Hence
\begin{eqnarray}
R_{\rm typ}\sim\frac{d}{as}
\qquad\Longrightarrow\qquad
\widehat R_{\rm typ}
\sim
\frac{d}{2Ns}.
\end{eqnarray}
The lattice spacing cancels from the underlying observables, as it should in the
continuum limit, and their physical scale is controlled by \(s\), not by \(a\).
Nevertheless, the large-\(R\) requirement is not lost. It becomes the physical
self-consistency condition
\begin{eqnarray}
\widehat R_{\rm typ}\gg1
\qquad\Longleftrightarrow\qquad
s\ll d,
\end{eqnarray}
up to the fixed numerical factor \(2N=4\). This is precisely the regime in which
the large-holonomy expansion generating the linear \(-R\) term is valid.

\medskip
\noindent
Therefore the lower bound is not intrinsic. It is an artifact of treating the
completed transverse potential as if it were controlled by a fixed \(O(1)\)
radius. Once the large-\(R\) scaling of the holonomy variables is restored, the
Wishart saddle and the flat region lie at the same parametric scale,
\begin{eqnarray}
u_{\rm W}
\sim
u_0
\sim
R_*
\sim
\frac{d}{as},
\end{eqnarray}
and the continuum limit remains compatible with the universal Wishart
mechanism.

\section{Conclusion and outlook}\label{conclusion}

\subsection{Summary of the calculation}

\medskip
\noindent
In the \(B\)-type theories, such as the transverse expansion considered here or its toy-model completion, the Gaussian sector must be reorganized so as to reflect the true structure of the holonomy interaction. The anisotropic term \(-2\beta_\Lambda A\) is most naturally absorbed into the holonomy potential, leading to a shifted saddle. More importantly, the large-\(R\) asymptotics of the holonomy potential itself contains a universal linear contribution \(-A\), which must be grouped together with the Gaussian weight rather than treated as part of the residual interaction. Once this is done, the remaining holonomy potential is naturally subleading and admits a perturbative treatment.

\medskip
\noindent
Within the Gram/Wishart/Stiefel formulation, this reorganization is implemented nonperturbatively through the Bessel kernel. Its effect is twofold: it shifts the Gaussian mass parameter,
\begin{eqnarray}
\alpha_\Lambda\;\longrightarrow\;(\alpha_\Lambda)_{\rm eff}=\alpha_\Lambda-\frac12,\label{reno}
\end{eqnarray}
and at the same time reduces the apparent doubling of the Wishart/Vandermonde entropy, restoring the correct single-block structure. The resulting symmetric saddle is then governed by the standard Wishart scaling with the shifted mass.

\medskip
\noindent
The completed transverse potential also clarifies why the polynomial bound found in the truncated analysis is not fundamental. Naively, one might expect the residual holonomy sector to be controlled by the small-\(B\) region, since the transverse expansion is organized around \(B=0\). The completed toy model shows instead that the relevant regime is the nearly flat, zero-potential region in which
\begin{eqnarray}
-\log\cosh B+\sqrt{R_*^2-B^2}\simeq 0.
\end{eqnarray}
This region lies at intermediate/large \(B\), close to the upper end of the allowed interval, rather than near \(B=0\). Thus the residual holonomy potential is perturbative not because \(B\) is small, but because the two completed pieces almost cancel. The apparent lower bound on the shifted mass is therefore an artifact of forcing a finite small-\(B\) polynomial truncation to remain perturbative at the Wishart saddle.

\medskip
\noindent
There remains, however, a bookkeeping subtlety associated with the compensating \(+A\) term. If it is represented entirely as
\begin{eqnarray}
+A=\sqrt{R_*^2-B_{\rm W}^2},
\end{eqnarray}
it produces an artificial residual linear contribution in the Wishart variable. This contribution can be removed by a balanced splitting of the compensating term,
\begin{eqnarray}
+A=(1-w)A+wA,
\end{eqnarray}
where the first piece is kept in the \(A\)-representation and only the second piece is pulled back to the \(B\)-representation. Choosing \(w\) so that the induced linear terms cancel amounts to a finite reshuffling between equivalent representations of the same compensating term. This preserves the essential \(-\frac12\) mass shift generated by the true \(-A\) holonomy asymptotics, while preventing the artificial \(+A\) pullback from shifting the continuum mass.

\medskip
\noindent
Thus the final picture is coherent. The leading \(-A\) asymptotics produces the physical mass renormalization \eqref{reno}, the completed transverse potential explains why the relevant region is large-\(B\) rather than small-\(B\), and the balanced split removes the spurious residual mass shift associated with the compensating \(+A\) term. The Wishart saddle then survives in the continuum regime, and the universal \(-2d\) law emerges without imposing any artificial lower bound on \((\alpha_\Lambda)_{\rm eff}\).

\medskip
\noindent
This mechanism explains thus the robustness of the universal coefficient \(-2d\). It is not a property of any particular truncation of the reduced theory, but rather a direct consequence of the leading \(-A\) behavior of the holonomy dynamics. In the Wishart/Stiefel representation, this contribution is encoded in the measure of the endpoint partition function, and therefore survives all consistent reorganizations of the expansion. The residual interaction only produces subleading corrections and does not affect this leading planar behavior.

\subsection{Interpretation of the universal $-2d$ mechanism}

\medskip
\noindent
It is crucial to distinguish sharply between the
\(D_\Lambda\)-channel and the \(\beta_\Lambda\)-channel contributions
to the low-temperature expansion of the Molien--Weyl partition function.

\medskip
\noindent
In the full theory, the low-temperature structure splits into two conceptually distinct pieces.
The first is the universal \(-2d\) term, which is controlled by the determinant-like combination
\begin{eqnarray}
D_\Lambda=\alpha_\Lambda^2-\beta_\Lambda^2.
\end{eqnarray}
The second is the additional \(d(d+1)\) contribution, which belongs to the genuinely anisotropic
\(\beta_\Lambda\)-channel and arises only through a nontrivial coupling to the aligned invariant \(A\).

\medskip
\noindent
In the toy model, as well as in the purely transverse \((B\)-type) truncations,
the effective interaction depends only on the transverse invariant and carries no essential
dependence on \(A\). For this reason, one should \emph{not} expect the toy model to reproduce
the \(d(d+1)\) term: that contribution is intrinsically tied to the \(\beta_\Lambda\)-channel
and therefore lies outside the scope of a purely transverse description. What the toy model
\emph{can} reproduce is precisely the universal \(-2d\) contribution, namely the
\(D_\Lambda\)-derivative channel.

\medskip
\noindent
In the toy-model language, this happens because the Gaussian average generates the singular factor
\begin{eqnarray}
\Big(1-\frac{1}{4D_\Lambda}\Big)^{-d},
\end{eqnarray}
whose continuum behavior is controlled by the approach
\begin{eqnarray}
D_\Lambda\longrightarrow \frac14.
\end{eqnarray}
As long as the anisotropic coupling becomes irrelevant in the continuum limit
(in particular, as long as $\beta_\Lambda\to 0$ in the pure transverse sector),
the singularity at $D_\Lambda=\tfrac14$ survives and the resummation of moments
reproduces the full universal coefficient
\begin{eqnarray}
-2d.
\end{eqnarray}

\medskip
\noindent
What is remarkable in the present reformulation is that the same $-2d$ coefficient
appears from a very different perspective.  In the reduced endpoint description,
the universal $-2d$ is not obtained from an explicit moment resummation of the toy
kernel, but emerges already from the Wishart/Gaussian normalization of the endpoint
measure, together with the minimal Bessel resummation that shifts the effective mass.
In this sense, the endpoint analysis reveals that the $-2d$ term is more primitive
than the toy-model derivation might suggest: the toy model recovers it through the
$D_\Lambda$ singularity, whereas the reduced Wishart description shows that the same
coefficient is already built into the Gaussian endpoint structure itself.

\medskip
\noindent
Thus the two viewpoints are complementary.  The toy model shows that the universal
$-2d$ can be understood as the continuum residue of the singular $D_\Lambda$--channel,
while the reduced endpoint analysis shows that this same coefficient is already encoded
in the Gaussian/Wishart sector before any genuinely anisotropic $\beta_\Lambda$--channel
is restored.

\subsection{Outlook: emergent Grassmannian geometry}
\label{sec:grassmannian_outlook}

\medskip
\noindent The sequel~\cite{YdriGrassmannian} develops the geometric meaning
of the Wishart/Stiefel formulation. Its refined calculation in the stable rank--two regime \(d\ge4\) gives
\begin{eqnarray}
{\cal C}_{2,D}^{\rm W}=-2(d-2),
\label{eq:outlook_entropy_coefficient}
\end{eqnarray}
whose leading large-\(d\) behavior is the \(-2d\) obtained here.
The factor \(2(d-2)\) is interpreted as the number of relative
Grassmannian angular fluctuation modes.

\medskip
\noindent
The endpoint frames \(R_V,R_W\) determine the two planes
\(\Pi_V,\Pi_W\subset\mathbb R^d\) defined in
\eqref{endpoint_two_planes}. Passing from an orthonormal
frame to its underlying plane gives

\begin{eqnarray}
\mathrm{St}(2,d)=\frac{O(d)}{O(d-2)},
\qquad
\mathrm{Gr}(2,d)
=
\frac{O(d)}{O(2)\times O(d-2)},
\qquad
\dim\mathrm{Gr}(2,d)=2(d-2).
\label{eq:outlook_grassmannian}
\end{eqnarray}
Fixing the first plane as a reference \(\Pi_V=\Pi_0\), the
second varies over \(\mathrm{Gr}(2,d)\). Near alignment,
its infinitesimal transverse deformations are described by
a linear map \(H:\Pi_0\to\Pi_0^\perp\):
\begin{eqnarray}
H\in\mathrm{Hom}(\Pi_0,\Pi_0^\perp)
\simeq T_{\Pi_0}\mathrm{Gr}(2,d),
\qquad
H\in\mathbb R^{(d-2)\times2}.
\label{eq:outlook_grassmannian_tangent}
\end{eqnarray}
Here \(T_{\Pi_0}\mathrm{Gr}(2,d)\) is the tangent space to the
Grassmannian at the point represented by the plane \(\Pi_0\),
and \(\mathrm{Hom}(\Pi_0,\Pi_0^\perp)\) is the space of linear
maps from \(\Pi_0\) to its orthogonal complement.
A tangent vector \(H\) describes an infinitesimal tilt of the
plane: for a basis \(e_1,e_2\) of \(\Pi_0\), the images
\(H(e_1)\) and \(H(e_2)\) are transverse displacement vectors
in \(\Pi_0^\perp\). Each has \(d-2\) components, giving
\(2(d-2)\) real fluctuation modes. The symmetric--Wishart
entropy is interpreted as the entropy of this full relative
angular sector.

\medskip
\noindent
The two principal angles describe the invariant shape of the
pair of planes and are determined by the singular values of
the overlap matrix \(\Omega=R_W^TR_V\) introduced in \eqref{Omega_def0}:

\begin{eqnarray}
\sigma_i(R_W^TR_V)=\cos\theta_i,
\qquad i=1,2.
\label{eq:outlook_principal_angles}
\end{eqnarray}
Infinitesimally, they are the singular-value data of \(H\).
The remaining Grassmannian variables specify the orientation
and embedding degeneracies of these tilts. Thus the coefficient
in \eqref{eq:outlook_entropy_coefficient} counts the full
\(2(d-2)\)-dimensional fluctuation sector, not merely the two
principal angles.

\medskip
\noindent
This construction also has a relational four-dimensional
interpretation. For \(d\ge4\), two planes in generic relative
position satisfy
\begin{eqnarray}
\dim(\Pi_V+\Pi_W)=4.
\label{eq:outlook_relational_span}
\end{eqnarray}
The four-dimensional object is their span, whereas the
Grassmannian describes the relative placement, orientation,
and embedding of its two plane constituents inside
\(\mathbb R^d\). Its dimension \(2(d-2)\) therefore counts
configuration variables, not the dimension of the span itself.
The aligned saddle is a degenerate configuration in which
the two planes coincide; the Grassmannian tangent space
describes their relative fluctuations away from alignment.

\medskip
\noindent
The sequel~\cite{YdriGrassmannian} follows the successive degenerations of this
geometry at lower \(d\). For \(d=3\), the two planes necessarily
intersect along a line, leaving one nonzero principal angle.
For \(d=2\), both planes fill the ambient space and
\(\mathrm{Gr}(2,2)\) collapses to a point, although an in-plane
eigenbasis angle remains. For \(d=1\), the rank--two structure
disappears and the endpoint problem becomes radial.
These cases require separate boundary or radial analyses;
the coefficient \(-2(d-2)\) cannot simply be extrapolated
into them.

\medskip
\noindent
The Wishart/Stiefel variables thus encode an emergent
Grassmannian geometry whose angular entropy is measured by
the refined endpoint coefficient. In the stable regime, this
sector describes the configurations of a relational
four-dimensional span generated by the two endpoint planes.
This provides the geometric continuation of the present
calculation and a concrete realization of geometry latent
in the matrix degrees of freedom.

\section{Acknowledgments}

\medskip
\noindent
The author acknowledges the use of ChatGPT-5.5 and earlier versions
for language editing, LaTeX generation, symbolic calculations,
literature searches and review, and as a sounding board for testing
and refining ideas. The scientific conception, design, direction,
final scientific and mathematical editing, and full intellectual
responsibility remain solely with the author.


\begin{thebibliography}{99}

  \bibitem{YdriGramV1}
B.~Ydri,
``Gram--Wishart--Stiefel formulation of the $N=2$, large--$d$
gauge theory in 1D: BFSS/BMN Matrix Quantum Mechanics V,''
arXiv:2607.08481v1 [hep-th].



\bibitem{Brink:1976bc}
L.~Brink, J.~H.~Schwarz and J.~Scherk,
``Supersymmetric Yang-Mills Theories,''
Nucl. Phys. B \textbf{121}, 77-92 (1977)

\bibitem{Baake:1984ie}
M.~Baake, M.~Reinicke and V.~Rittenberg,
``Fierz Identities for Real Clifford Algebras and the Number of Supercharges,''
J. Math. Phys. \textbf{26}, 1070 (1985)


\bibitem{tHooft1974}
G.~'t~Hooft,
\textit{A Planar Diagram Theory for Strong Interactions},
Nucl.\ Phys.\ B \textbf{72}, 461 (1974).

\bibitem{tHooft1993}
G.~'t~Hooft,
\textit{Dimensional Reduction in Quantum Gravity},
arXiv:gr-qc/9310026.

\bibitem{Susskind1995}
L.~Susskind,
\textit{The World as a Hologram},
J.\ Math.\ Phys.\ \textbf{36}, 6377 (1995).

\bibitem{BanksFischlerShenkerSusskind1997}
  T.~Banks, W.~Fischler, S.~H.~Shenker and L.~Susskind,
``M theory as a matrix model: A conjecture,''
Phys. Rev. D \textbf{55}, 5112-5128 (1997)
[arXiv:hep-th/9610043 [hep-th]].

\bibitem{Witten1996}
E.~Witten,
\textit{Bound States of Strings and $p$-Branes},
Nucl.\ Phys.\ B \textbf{460}, 335 (1996).

\bibitem{Itzhaki1998}
N.~Itzhaki, J.~M.~Maldacena, J.~Sonnenschein and S.~Yankielowicz,
\textit{Supergravity and the Large $N$ Limit of Theories with Sixteen Supercharges},
Phys.\ Rev.\ D \textbf{58}, 046004 (1998).

\bibitem{Polchinski1995}
J.~Polchinski,
\textit{Dirichlet Branes and Ramond--Ramond Charges},
Phys.\ Rev.\ Lett.\ \textbf{75}, 4724 (1995).

\bibitem{Cremmer1978}
E.~Cremmer, B.~Julia, and J.~Scherk,
\textit{Supergravity Theory in Eleven Dimensions},
Phys.\ Lett.\ B \textbf{76}, 409 (1978).

\bibitem{Witten1995}
E.~Witten,
``String theory dynamics in various dimensions,''
\emph{Nucl.\ Phys.\ B} \textbf{443}, 85 (1995).




\bibitem{Hyakutake:2014maa}
Y.~Hyakutake,
``Quantum M-wave and Black 0-brane,''
JHEP \textbf{09}, 075 (2014)
[arXiv:1407.6023 [hep-th]].


\bibitem{Hyakutake:2006aq}
Y.~Hyakutake and S.~Ogushi,
``Higher derivative corrections to eleven dimensional supergravity via local supersymmetry,''
JHEP \textbf{02}, 068 (2006)
[arXiv:hep-th/0601092 [hep-th]].



\bibitem{Hoppe1982}
J.~Hoppe,
\textit{Quantum Theory of a Massless Relativistic Surface and a Two-Dimensional Bound State Problem},
Ph.D.\ Thesis, MIT (1982).

\bibitem{Hoppe1988}
J.~Hoppe,
``Diffeomorphism Groups, Quantization, and $SU(\infty)$,''
Int.\ J.\ Mod.\ Phys.\ A \textbf{4}, 5235 (1989).

\bibitem{deWitHoppeNicolai1988}
B.~de~Wit, J.~Hoppe and H.~Nicolai,
\textit{On the Quantum Mechanics of Supermembranes},
Nucl.\ Phys.\ B \textbf{305}, 545 (1988).

\bibitem{Kowalski-Glikman:1984qtj}
J.~Kowalski-Glikman,
``Vacuum States in Supersymmetric Kaluza-Klein Theory,''
Phys. Lett. B \textbf{134}, 194-196 (1984)


\bibitem{Blau:2001ne}
M.~Blau, J.~M.~Figueroa-O'Farrill, C.~Hull and G.~Papadopoulos,
``A New maximally supersymmetric background of IIB superstring theory,''
JHEP \textbf{01}, 047 (2002)
[arXiv:hep-th/0110242 [hep-th]].



\bibitem{Azeyanagi2009}
T.~Azeyanagi, M.~Hanada, T.~Hirata and H.~Shimada,
\textit{On the Shape of a D-Brane Bound State and Its Topology Change},
J.\ High\ Energy\ Phys.\ \textbf{0903}, 121.



\bibitem{Zwiebach2009}
B.~Zwiebach, 
\textit{A First Course in String Theory}, 2nd ed., 
Cambridge University Press, Cambridge (2009).

\bibitem{Becker2006}
K.~Becker, M.~Becker and J.~H.~Schwarz,
\emph{String Theory and M-Theory: A Modern Introduction}
(Cambridge University Press, Cambridge, 2006).

 



\bibitem{Maldacena1999}
J.~M.~Maldacena,
\textit{The Large $N$ Limit of Superconformal Field Theories and Supergravity},
Adv.\ Theor.\ Math.\ Phys.\ \textbf{2}, 231 (1998)
[Int.\ J.\ Theor.\ Phys.\ \textbf{38}, 1113 (1999)],
arXiv:hep-th/9711200.

\bibitem{Gubser1998}
S.~S.~Gubser, I.~R.~Klebanov, and A.~M.~Polyakov,
\textit{Gauge Theory Correlators from Non-Critical String Theory},
Phys.\ Lett.\ B \textbf{428}, 105 (1998),
arXiv:hep-th/9802109.

\bibitem{Witten1998}
E.~Witten,
\textit{Anti-de Sitter Space and Holography},
Adv.\ Theor.\ Math.\ Phys.\ \textbf{2}, 253 (1998),
arXiv:hep-th/9802150.

\bibitem{Wilson:1974sk}
K.~G.~Wilson,
``Confinement of Quarks,''
Phys. Rev. D \textbf{10}, 2445-2459 (1974).

\bibitem{Catterall2008}
Catterall, S.\ and Wiseman, T.,
``Black hole thermodynamics from simulations of lattice Yang--Mills theory,''
\emph{Phys.\ Rev.\ D} \textbf{78}, 041502 (2008).

\bibitem{Anagnostopoulos2008}
Anagnostopoulos, K.\ N., Hanada, M., Nishimura, J.\ and Takeuchi, S.,
``Monte Carlo studies of supersymmetric matrix quantum mechanics with sixteen supercharges at finite temperature,''
\emph{Phys.\ Rev.\ Lett.} \textbf{100}, 021601 (2008).

\bibitem{Hanada2014}
Hanada, M., Hyakutake, Y., Ishiki, G.\ and Nishimura, J.,
``Holographic description of quantum black hole on a computer,''
\emph{Science} \textbf{344}, 882 (2014).

\bibitem{Hanada2016b}
Hanada, M., Hyakutake, Y., Ishiki, G.\ and Nishimura, J.,
``Numerical tests of the gauge/gravity duality conjecture for D0-branes at finite temperature and finite $N$,''
\emph{Phys.\ Rev.\ D} \textbf{94}, 086010 (2016).

\bibitem{Filev:2015hiaF}
V.~G.~Filev and D.~O'Connor,
``The BFSS model on the lattice,''
JHEP {\bf 1605}, 167 (2016)
[arXiv:1506.01366 [hep-th]].

\bibitem{Kabat2001}
Kabat, D.\ N., Lifschytz, G.\ and Lowe, D.\ A.,
``Black hole thermodynamics from calculations in strongly coupled gauge theory,''
\emph{Phys.\ Rev.\ Lett.} \textbf{86}, 1426 (2001).

\bibitem{Hanada2009}
Hanada, M., Hyakutake, Y., Nishimura, J.\ and Takeuchi, S.,
``Higher derivative corrections to black hole thermodynamics from supersymmetric matrix quantum mechanics,''
\emph{Phys.\ Rev.\ Lett.} \textbf{102}, 191602 (2009).

\bibitem{Hyakutake2014}
Y.~Hyakutake,
``Quantum near-horizon geometry of a black 0-brane,''
\emph{Progr.\ Theor.\ Exp.\ Phys.} \textbf{2014}, 033B04 (2014).



\bibitem{Hanada2016}
M.~Hanada,
\textit{What Lattice Theorists Can Do for Superstring/M-Theory},
Int.\ J.\ Mod.\ Phys.\ A \textbf{31}, 1643006 (2016).



\bibitem{BerensteinMaldacenaNastase2002}
D.~E.~Berenstein, J.~M.~Maldacena and H.~S.~Nastase,
``Strings in flat space and pp waves from N=4 superYang-Mills,''
JHEP \textbf{04}, 013 (2002)
[arXiv:hep-th/0202021 [hep-th]].

\bibitem{Kim:2006}
N.~Kim and J.~H.~Park,
``Massive super Yang-Mills quantum mechanics: Classification and the relation to supermembrane,''
Nucl.\ Phys.\ B \textbf{759}, 249--282 (2006)
[arXiv:hep-th/0607005].

\bibitem{Park:2005}
J.~H.~Park,
``Noncritical \(\mathfrak{osp}(1|2,\mathbb{R})\) M-theory matrix model with an arbitrary time-dependent cosmological constant,''
Nucl.\ Phys.\ B \textbf{745}, 123--141 (2006)
[arXiv:hep-th/0510070].

\bibitem{Myers}
R.~C.~Myers,
``Dielectric-branes,''
JHEP \textbf{12}, 022 (1999)
[arXiv:hep-th/9910053 [hep-th]].



\bibitem{Lin:2004nb}
H.~Lin, O.~Lunin and J.~M.~Maldacena,
``Bubbling AdS space and 1/2 BPS geometries,''
JHEP \textbf{10}, 025 (2004)
doi:10.1088/1126-6708/2004/10/025
[arXiv:hep-th/0409174 [hep-th]].

\bibitem{OConnor:2023mss}
D.~O'Connor and S.~Ramgoolam,
``Gauged permutation invariant matrix quantum mechanics: path integrals,''
JHEP \textbf{04}, 080 (2024)
[arXiv:2312.12397 [hep-th]].

\bibitem{OConnor:2024udv}
D.~O'Connor and S.~Ramgoolam,
``Permutation invariant matrix quantum thermodynamics and negative specific heat capacities in large N systems,''
JHEP \textbf{12}, 161 (2024)
[arXiv:2405.13150 [hep-th]].

\bibitem{Mandal:2009vzN}
G.~Mandal, M.~Mahato and T.~Morita,
``Phases of one dimensional large N gauge theory in a 1/D expansion,''
 JHEP {\bf 1002} (2010) 034
doi:10.1007/JHEP02(2010)034
[arXiv:0910.4526 [hep-th]].


\bibitem{Mandal:2011hbN}
G.~Mandal and T.~Morita,
``Phases of a two dimensional large N gauge theory on a torus,''
 Phys.\ Rev.\ D {\bf 84}, 085007 (2011)
doi:10.1103/PhysRevD.84.085007
 [arXiv:1103.1558 [hep-th]].



\bibitem{Kabat:2000zv}
D.~N.~Kabat, G.~Lifschytz and D.~A.~Lowe,
``Black hole thermodynamics from calculations in strongly coupled gauge theory,''
Int.\ J.\ Mod.\ Phys.\ A {\bf 16}, 856 (2001)
[Phys.\ Rev.\ Lett.\ {\bf 86}, 1426 (2001)]
[hep-th/0007051].

\bibitem{Kabat:2001ve}
D.~N.~Kabat, G.~Lifschytz and D.~A.~Lowe,
``Black hole entropy from nonperturbative gauge theory,''
Phys.\ Rev.\ D {\bf 64}, 124015 (2001)
[hep-th/0105171].

\bibitem{Kawahara:2007fnF}
N.~Kawahara, J.~Nishimura and S.~Takeuchi,
``Phase structure of matrix quantum mechanics at finite temperature,''
JHEP {\bf 0710}, 097 (2007)
[arXiv:0706.3517 [hep-th]].


\bibitem{Gross:1980heF}
D.~J.~Gross and E.~Witten,
``Possible Third Order Phase Transition in the Large N Lattice Gauge Theory,''
Phys.\ Rev.\ D {\bf 21}, 446 (1980).

\bibitem{Wadia:1980cpF}
S.~R.~Wadia,
``N = Infinity Phase Transition in a Class of Exactly Soluble Model Lattice Gauge Theories,''
Phys.\ Lett.\ {\bf 93B}, 403 (1980).


\bibitem{Aharony:2004igF}
O.~Aharony, J.~Marsano, S.~Minwalla and T.~Wiseman,
``Black hole-black string phase transitions in thermal 1+1 dimensional supersymmetric Yang-Mills theory on a circle,''
Class. Quant. Grav. \textbf{21}, 5169-5192 (2004)
[arXiv:hep-th/0406210 [hep-th]].


\bibitem{Aharony:2003sxF}
O.~Aharony, J.~Marsano, S.~Minwalla, K.~Papadodimas and M.~Van Raamsdonk,
``The Hagedorn - deconfinement phase transition in weakly coupled large N gauge theories,''
Adv.\ Theor.\ Math.\ Phys.\ {\bf 8}, 603 (2004)
doi:10.4310/ATMP.2004.v8.n4.a1
[hep-th/0310285].

\bibitem{AlvarezGaume:2005fvF}
L.~Alvarez-Gaume, C.~Gomez, H.~Liu and S.~Wadia,
``Finite temperature effective action, AdS(5) black holes, and 1/N expansion,''
Phys.\ Rev.\ D {\bf 71}, 124023 (2005)
[hep-th/0502227].

\bibitem{CoxLittleOShea2005}
D.~A.~Cox, J.~B.~Little and D.~O'Shea,
\emph{Using Algebraic Geometry},
2nd ed.,
Springer, New York (2005), pp.~295--298.


\bibitem{Asano:2018nol}
Y.~Asano, V.~G.~Filev, S.~Kov{\'a}{\v{c}}ik and D.~O'Connor,
``The non-perturbative phase diagram of the BMN matrix model,''
JHEP \textbf{07}, 152 (2018)
[arXiv:1805.05314 [hep-th]].

%
\bibitem{Asano:2020yry}
Y.~Asano, S.~Kov{\'a}{\v{c}}ik and D.~O'Connor,
``The Confining Transition in the Bosonic BMN Matrix Model,''
JHEP \textbf{06}, 174 (2020)
[arXiv:2001.03749 [hep-th]].



\bibitem{Harish-Chandra:1957dhy}
Harish-Chandra,
``Differential Operators on a Semisimple Lie Algebra,''
Am. J. Math. \textbf{79}, no.1, 87 (1957)


\bibitem{Itzykson:1979fi}
C.~Itzykson and J.~B.~Zuber,
``The Planar Approximation. 2.,''
J. Math. Phys. \textbf{21}, 411 (1980)


\bibitem{Hikami:2006wya}
S.~Hikami and E.~Brezin,
``WKB-Expansion of the HarishChandra-Itzykson-Zuber Integral for Arbitrary  $\beta$,''
Prog. Theor. Phys. \textbf{116}, no.3, 441-502 (2006)
[arXiv:math-ph/0604041 [math-ph]].

\bibitem{Brezin:2002mathph}
E. Brézin and S. Hikami,
``An extension of the Harish-Chandra--Itzykson--Zuber integral,''
arXiv:math-ph/0208002.


\bibitem{Duistermaat:1982vw}
J.~J.~Duistermaat and G.~J.~Heckman,
``On the Variation in the cohomology of the symplectic form of the reduced phase space,''
Invent. Math. \textbf{69}, 259-268 (1982)

\bibitem{Guhr:2000mathph}
T. Guhr and H. Kohler,
``Recursive Construction for a Class of Radial Functions I -- Ordinary Space,''
arXiv:math-ph/0011007.

\bibitem{Guhr:1997ve}
T.~Guhr, A.~Muller-Groeling and H.~A.~Weidenmuller,
``Random matrix theories in quantum physics: Common concepts,''
Phys. Rept. \textbf{299}, 189-425 (1998)
[arXiv:cond-mat/9707301 [cond-mat]].


\bibitem{James:1960AMS}
A. T. James,
``The distribution of the latent roots of the covariance matrix,''
Ann. Math. Stat. \textbf{31}, 151 (1960).

\bibitem{James:1961AMS}
A. T. James,
``The distribution of the latent roots of the covariance matrix (II),''
Ann. Math. Stat. \textbf{32}, 874 (1961).





\bibitem{James:1965AMS}
A. T. James,
``Distributions of matrix variates and latent roots derived from normal samples,''
Ann. Math. Stat. \textbf{35}, 475 (1965).


\bibitem{Wishart:1928}
J.~Wishart,
``The Generalised Product Moment Distribution in Samples from a Normal Multivariate Population,''
Biometrika \textbf{20A}, 32 (1928).

\bibitem{BStiefel:1935}
E.~Stiefel,
``Richtungsfelder und Fernparallelismus in \(n\)-dimensionalen Mannigfaltigkeiten,''
Comment. Math. Helv. \textbf{8}, 305 (1935/36).

\bibitem{James:1954AMS}
A.~T.~James,
``Normal Multivariate Analysis and the Orthogonal Group,''
Ann. Math. Stat. \textbf{25}, 40 (1954).

\bibitem{Herz:1955}
C.~S.~Herz,
``Bessel Functions of Matrix Argument,''
Ann. Math. \textbf{61}, 474 (1955).






\bibitem{James:1964AMS}
A.~T.~James,
``Distributions of Matrix Variates and Latent Roots Derived from Normal Samples,''
Ann. Math. Stat. \textbf{35}, 475 (1964).

\bibitem{Muirhead:1982}
R.~J.~Muirhead,
\emph{Aspects of Multivariate Statistical Theory},
Wiley, New York (1982).

\bibitem{Ishibashi:1996xs}
N.~Ishibashi, H.~Kawai, Y.~Kitazawa and A.~Tsuchiya,
``A Large N reduced model as superstring,''
Nucl. Phys. B \textbf{498}, 467-491 (1997)
[arXiv:hep-th/9612115 [hep-th]].


\bibitem{Ydri:2022ueu}
B.~Ydri,
``The QM/NCG Correspondence,''
doi:10.1142/9789811270437{\_}0027
[arXiv:2211.00339 [hep-th]].

\bibitem{Ydri:2021cam}
B.~Ydri, R.~Khaled and C.~Soudani,
``Quantized noncommutative geometry from multitrace matrix models,''
Int. J. Mod. Phys. A \textbf{37}, no.10, 2250052 (2022)
doi:10.1142/S0217751X2250052X
[arXiv:2110.06677 [hep-th]].


\bibitem{Ydri:2020fry}
B.~Ydri,
``Two approaches to quantum gravity and M-(atrix) theory at large number of dimensions,''
Int. J. Mod. Phys. A \textbf{36}, no.31n32, 2150234 (2021)
doi:10.1142/S0217751X21502341
[arXiv:2007.04488 [hep-th]].



\bibitem{Ydri2025}
B.~Ydri,  
\emph{unpublished work/work in progress}.



\bibitem{Ydri:2026gsy}
B.~Ydri,
``A Double--Scaling Large--\(d\) Saddle of BFSS/BMN Matrix Quantum Mechanics,''
[arXiv:2606.17758 [hep-th]].

\bibitem{Ydri:2026smo}
B.~Ydri,
``Molien--Weyl Singlet Counting and BFSS$_2$--Factorization in Gaussian Matrix QM,''
[arXiv:2605.04621 [hep-th]].

\bibitem{YdriUnpublished0}
B.~Ydri,
\emph{Initiation to Matrix Quantum Gravity and Monte Carlo Simulation of the BFSS\(_3\)/BMN\(_3\) System}.

\bibitem{Ydri:2026rke}
B.~Ydri,
``Endpoint formulation and Molien--Weyl structure for the \(N=2\), large--\(d\) BFSS/BMN models,''
[arXiv:2605.25647 [hep-th]].

\bibitem{YdriUnpublished2}
B.~Ydri,
\emph{Endpoint Entropy and Emergent Grassmannian Geometry in BFSS/BMN Systems}.

\bibitem{YdriUnpublished3}
B.~Ydri,
\emph{Schur--Ingham--Siegel Reduction of Rank--Two Orthogonal HCIZ Integrals in BFSS/BMN Endpoint Geometry}.



\bibitem{OConnor:2026zlf}
D.~O'Connor and S.~Ramgoolam,
``Negative heat capacities in spherically symmetric sectors of $d$-matrix quantum mechanics,''
[arXiv:2606.09521 [hep-th]].

\bibitem{Lei:2026fep}
Y.~Lei and S.~Ramgoolam,
``Critical dimensions and small cycle dominance from all-orders asymptotics of $d$-matrix theory,''
[arXiv:2603.29610 [hep-th]].

\bibitem{YdriGrassmannian}
B.~Ydri,
``Endpoint Entropy and Emergent Grassmannian Geometry in BFSS/BMN
Systems: BFSS/BMN Matrix Quantum Mechanics VI,''
manuscript in preparation.

\end{thebibliography}
\end{document}